\documentclass[12pt]{article}
\usepackage{newtxtext,newtxmath}

\usepackage{graphicx}

\usepackage[letterpaper,margin=1in]{geometry}
\usepackage{authblk}

\renewenvironment{abstract}
	{\quotation}
	{\endquotation}

\date{}

\makeatletter
\renewcommand{\fnum@figure}{\textbf{Figure \thefigure}}
\renewcommand{\fnum@table}{\textbf{Table \thetable}}
\makeatother

\usepackage{cite}

\usepackage{url}

\usepackage{xcolor}
\usepackage[colorlinks=true,linkcolor=blue!60!black,
  citecolor=blue!60!black,urlcolor=blue!60!black,
  hypertexnames=false]{hyperref}

\newcommand{\br}{\mathbf{r}}
\newcommand{\bv}{\mathbf{v}}
\newcommand{\vna}{v{_\mathrm{Na}}}
\newcommand{\rna}{r_{\mathrm{Na}}}
\newcommand{\angmom}{\langle L_x \rangle}
\newcommand{\mps}{\mathrm{m\,s}^{-1}}

\def\scititle{
	Critical velocity and dissipation mechanism in finite superfluid $^4$He nanodroplets
}
\title{\bfseries \boldmath \scititle}

\author[1,2,*]{Keshav Sishodia}
\author[3]{Ernesto Garc{\'i}a-Alfonso}
\author[3]{Nadine Halberstadt}
\author[1]{Sivarama Krishnan}
\author[4,**]{Marcel Mudrich}

\affil[1]{Quantum Center of Excellence for Diamond and Emergent Materials, School of Interdisciplinary Sciences and Department of Physics, Indian Institute of Technology Madras, Chennai, India}
\affil[2]{ELI Beamlines facility, The Extreme Light Infrastructure ERIC, Doln{\'i} B\v{r}e\v{z}any, Czech Republic}
\affil[3]{Laboratoire Collisions, Agr{\'e}gats, R{\'e}activit{\'e} (LCAR), Universit{\'e} de Toulouse, CNRS, Toulouse, France}
\affil[4]{Institute of Physics and Center for Interdisciplinary Nanostructure Science and Technology (CINSaT), University of Kassel, Kassel, Germany}
\affil[*]{Corresponding author. Email: keshav.sishodia@eli-laser.eu}
\affil[**]{Corresponding author. Email: mudrich@uni-kassel.de}

\begin{document}

\maketitle

\begin{abstract} \bfseries \boldmath

A hallmark of superfluidity is the frictionless motion of an object below the Landau critical velocity; Whether this concept applies to nanoscale droplets remains unclear. Using time-dependent density functional theory, we investigate the motion of a sodium atom on the surface and inside a superfluid helium nanodroplet. For surface motion, we identify a critical velocity above which energy dissipation through ripplon excitation and the formation of surface-bound quantized vortices leads to rapid deceleration. Inside the droplet, the atom is slowed down to a steady-state velocity that depends on the initial speed, contrasting experiments reporting a critical velocity for dopants ejected from droplets [Phys. Rev. Lett. 111, 153002 (2013)]. Remarkably, the velocity of the ejected atom is nearly independent of its initial velocity. We demonstrate that this behavior results from the bursting of the bubble surrounding the atom at the droplet surface, challenging the previous experimental interpretation.

\end{abstract}

\noindent
Superfluid helium (He-II) is unique among condensed-phase systems in exhibiting macroscopic quantum-fluid behavior. Its remarkable properties include exceptionally high thermal conductivity, vanishing viscosity for flow velocities below a critical value, irrotational flow, and the formation of quantized vortices when the superfluid is forced to carry angular momentum.
Since the initial discovery of superfluidity by Kapitza~\cite{Kapitza1938}, the properties of bulk He-II have been extensively investigated, both experimentally and theoretically—with Landau being the most prominent figure in the development of the theoretical framework~\cite{landauTheorySuperfluidityHelium1941}.
Nevertheless, the mechanisms governing the onset of dissipation, and the corresponding critical velocities associated with different excitation processes, remain subjects of active research~\cite{skrbek2025critical}.

New questions arise when the superfluid is confined to nanometer-sized droplets. To what extent do the characteristic properties of bulk He-II, such as frictionless flow, persist in finite systems where confinement and surface effects become increasingly important as the size decreases?
Does a critical velocity for frictionless motion of an impurity still exist in a He nanodroplet? How does the situation change when the impurity moves along the droplet surface rather than through its interior, where surface excitations, or ripplons, become relevant in addition to bulk excitations such as phonons and rotons? In a rotating nanodroplet, both ripplons and quantized vortices can carry angular momentum, raising the question of which excitation dominates at low velocities and how the two interact or potentially interconvert when simultaneously excited.

Freely propagating He nanodroplets offer several advantages over bulk He-II confined within a cryostat for investigating critical velocities. In particular, they provide a container-free environment, eliminating dissipation mechanisms associated with solid boundaries and reducing complications arising from remnant vortices~\cite{skrbek2025critical}. Moreover, He nanodroplets can be prepared with a high degree of chemical purity, minimizing the influence of impurities on superfluid dynamics, especially at the droplet surface. The only surface-bound contaminant that is difficult to eliminate, $^3$He, is typically present at a concentration of $\sim1$~ppm, such that the vast majority of He nanodroplets can be considered contaminant-free. These characteristics make He nanodroplets particularly attractive model systems for investigating the intrinsic mechanisms of energy dissipation, elementary excitations, and critical velocities in finite quantum fluids.

To address these questions,
we performed $^4$He time-dependent density functional theory ($^4$He-TDDFT) simulations of a He nanodroplet containing an impurity atom initially located either at the surface or in the interior. The atom is set in motion with a controlled initial velocity, and the ensuing coupled dynamics of the impurity and the superfluid are followed in real time. This work was motivated by recent experiments that revealed a surprisingly long timescale for the interatomic Coulombic decay (ICD) process involving two excited He atoms, He$^*$, that are weakly bound to the surface of a He nanodroplet~\cite{sishodiaXUVFluorescenceProbe2026}. The He$^*$ atoms appeared to move freely along the droplet surface without friction until they eventually collided with each other and decayed by ICD.

The sodium atom (Na) was chosen as an impurity because, like all alkali atoms, it is weakly bound in a dimple state at the surface of the droplet, thus serving as a viable probe of surface motion~\cite{stienkemeierSpectroscopyAlkaliAtoms1996, barrancoHeliumNanodropletsOverview2006}. In addition, it is sufficiently massive to be treated as a classical particle. Our initial goal was to investigate the existence of a critical velocity for surface motion, analogous to the Landau critical velocity in bulk He-II, $v_L\sim58~\mps$~\cite{landauTheorySuperfluidityHelium1941,mcclintock1995landau}.
This naturally led us to examine the motion of a Na atom artificially placed at the center of the He droplet to benchmark our method against previous simulation results~\cite{lealCaptureHeliophobicAtoms2014,vilaQuantumDynamicsPick2016,coppensHeadonCollisionsXe2017,coppensCaptureXeAr2017,garcia-alfonsoClusteringCollisionRelaxation2022,bonhommeauCollisionCesiumAtoms2024} and to critically assess the interpretation of an earlier experiment by Brauer et al.~\cite{brauerCriticalLandauVelocity2013}. In that experiment, dopant atoms or molecules initially embedded in He nanodroplets were accelerated out of the droplets by means of optical excitation, and the speed distributions of the ejected particles were measured. For all species, the resulting speeds were in the range of 40–$60~\mps$. This finding was taken as evidence for a critical velocity inside the nanodroplets. Accompanying TDDFT simulations seemed to support this interpretation, although the velocity of the molecules inside the nanodroplet remained significantly lower than $v_L$~\cite{brauerCriticalLandauVelocity2013}.

Further similar simulations
studied the motion of various types of atoms inside He nanodroplets, mainly from the perspective of the pickup process~\cite{lealCaptureHeliophobicAtoms2014,vilaQuantumDynamicsPick2016,coppensHeadonCollisionsXe2017,coppensCaptureXeAr2017,garcia-alfonsoClusteringCollisionRelaxation2022,bonhommeauCollisionCesiumAtoms2024}. In particular, the collision of an impurity particle with a He nanodroplet and its subsequent motion through the droplet were investigated for Cs, Ne, Ar, and Xe atoms~\cite{lealCaptureHeliophobicAtoms2014,vilaQuantumDynamicsPick2016,coppensHeadonCollisionsXe2017,coppensCaptureXeAr2017,garcia-alfonsoClusteringCollisionRelaxation2022,bonhommeauCollisionCesiumAtoms2024}.
However, the role of the Landau critical velocity was not explicitly discussed in those works, likely due to the lack of systematic simulation data.
The case closest to our current study is that of a Cs atom colliding with a He droplet at a relatively high velocity ($200~\mps$).
The Cs atom was slowed almost to a standstill inside the droplet~\cite{coppensCaptureXeAr2017} and eventually expelled from the droplet with a final speed of $64.7~\mps$~\cite{lealCaptureHeliophobicAtoms2014}, which happens to be close to $v_L$, but no claim was made regarding the existence of a critical velocity.
We will return to this case in the discussion of our results.

For peripheral collisions of an impurity atom with a He nanodroplet at a lower velocity ($50~\mps$), the impinging atom formed a dimple that rotated around the droplet, carrying the atom with it~\cite{lealCaptureHeliophobicAtoms2014}.
While the atom's velocity was significantly decreased in the process of attaching to the He nanodroplet and orbiting around it, its limiting speed and the possible existence of a critical velocity were not discussed in that work.

\section*{Results}

\subsection*{Surface motion}

\begin{figure}
\centering
\includegraphics[width=0.8\linewidth]{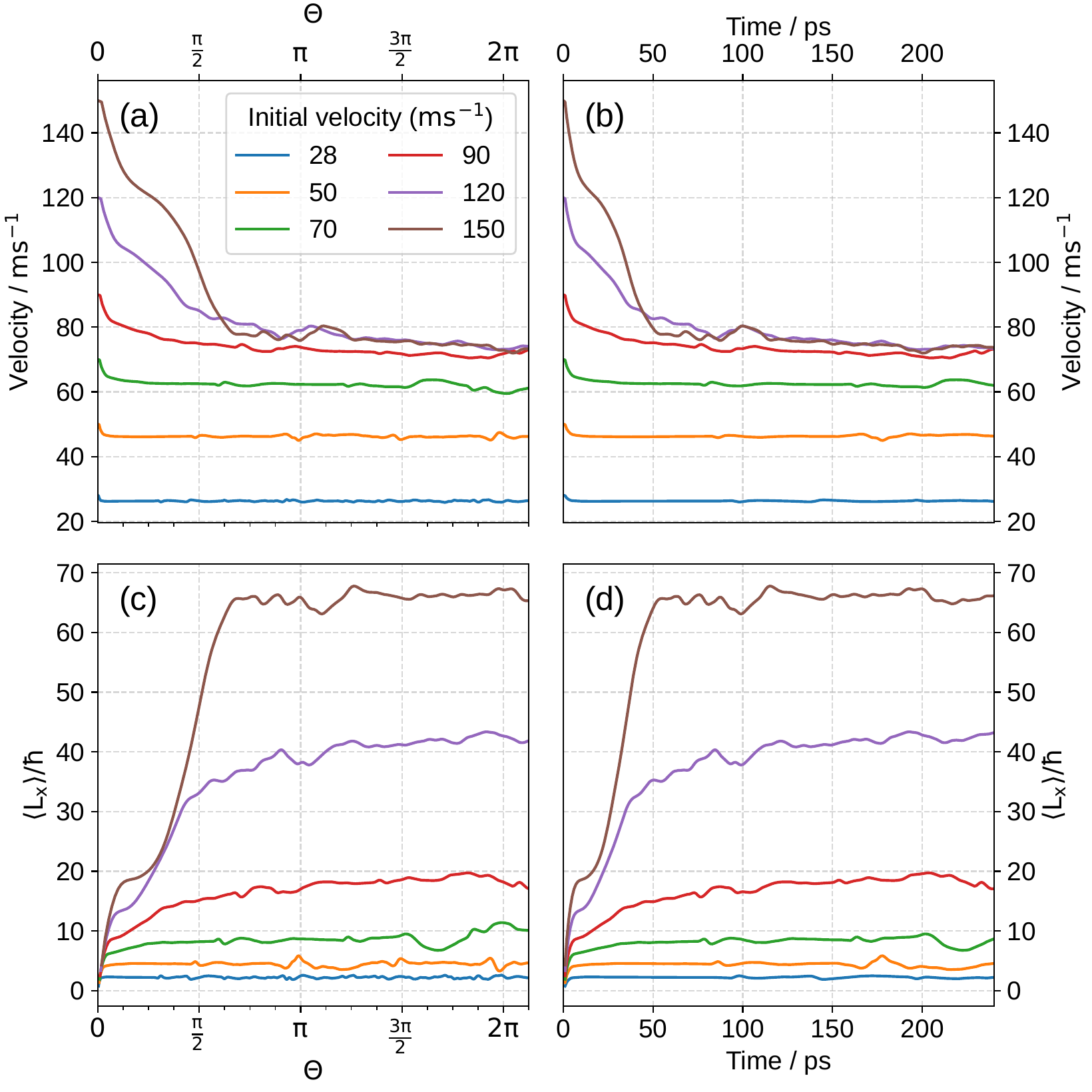}
\caption{
    \textbf{Motion of a Na impurity on the surface of $^4$He$_{1000}$ nanodroplets.}
    (a) Evolution of the speed of the Na impurity, $v_\mathrm{Na}$, as a function of the polar angle $\Theta$ (in spherical coordinates with the origin at the droplet center) and (b) as a function of time for different initial values of $\vna$ initially located at the surface of a He nanodroplet. (c) and (d) show the angular momentum $\angmom$ carried by the He density alone. For all values of the initial velocity, the Na atom is rapidly decelerated within $<10$~ps due to its collision with the dimple in which it resides. For initial speeds $\geq 70$~ms$^{-1}$, the Na atom further slows down to a final speed of $\sim70$~ms$^{-1}$ within $\sim 50$~ps while moving roughly a quarter turn around the droplet ($\Theta=\pi/2$). The concurrent increase in $\angmom$ indicates the transfer of angular momentum from the Na motion to the droplet.
    \label{fig:surf_comb}
}
\end{figure}

As a starting point for the TDDFT simulations, we computed the equilibrium configuration of a Na atom on the surface of a $^4$He$_{1000}$ nanodroplet (sharp-density radius $R=22.2$~\AA). For a detailed description of the method, see the review articles
\cite{barrancoHeliumNanodropletsOverview2006,
ancilottoDensityFunctionalTheory2017}.
At $t=0$, the Na atom resides in a surface dimple of radius $5.65~$\AA~at a distance from the droplet center of $25.65~$\AA, see Fig.~\ref{sfig:static}(a).
The simulations were initialized by assigning the Na atom a nonzero velocity parallel to the surface
in the $y$-direction, $\vna$.
See Materials and Methods for more details about the simulation
parameters. Fig.~\ref{fig:surf_comb} displays the velocity $\vna(t)$ of
the Na atom and the corresponding angular momentum $\langle L \rangle (t)$
of the He nanodroplet for selected initial velocities specified in (a).
For the chosen geometry (initial Na position on the $z$ axis and velocity $\vna$ along $y$), $\langle L \rangle (t)$ has a component only in the $x$-direction, $\angmom$, shown in Figs.~\ref{fig:surf_comb}~(c) and (d). Panel (a) shows the evolution of $\vna(t)$ as a function of
the polar angle  $\Theta$ of the Na atom on the droplet surface with respect to the droplet center, while (b) shows the same as a function of the simulation time.

For an initial Na velocity $\vna(t=0) \leq 70~\mps$ (red line),
the Na atom is only slightly decelerated within $t\lesssim 10$~ps as it collides with the dimple in which it initially resides.
This transient follows from assigning a velocity to the Na atom while the helium initially remains in its static configuration: the atom and its dimple must first adjust to collective motion.
It should therefore be distinguished from the subsequent damping used to identify the surface critical velocity;
see the Supplementary Material (SM) for an estimate of this initial slowdown based on the elastic-collision approximation for the Na atom and its surrounding dimple.
This initial loss of the atom's velocity is reflected in a gain in the
angular momentum of the droplet, $\angmom(t)$, as seen in Figs.~\ref{fig:surf_comb}(c)-(d). After the initial slowdown, the Na atom moves along the droplet surface with constant speed, indicating vanishing energy dissipation. When the initial Na speed is set to values $\vna > 70~\mps$, the dynamics changes qualitatively. In contrast to the cases where $\vna \leq 70~\mps$, the Na atom further slows to $\approx70~\mps$ over a period of $t \lesssim 10$~ps, while $\angmom$ increases correspondingly. 

\begin{figure}
\centering
\includegraphics[width=0.8\linewidth]{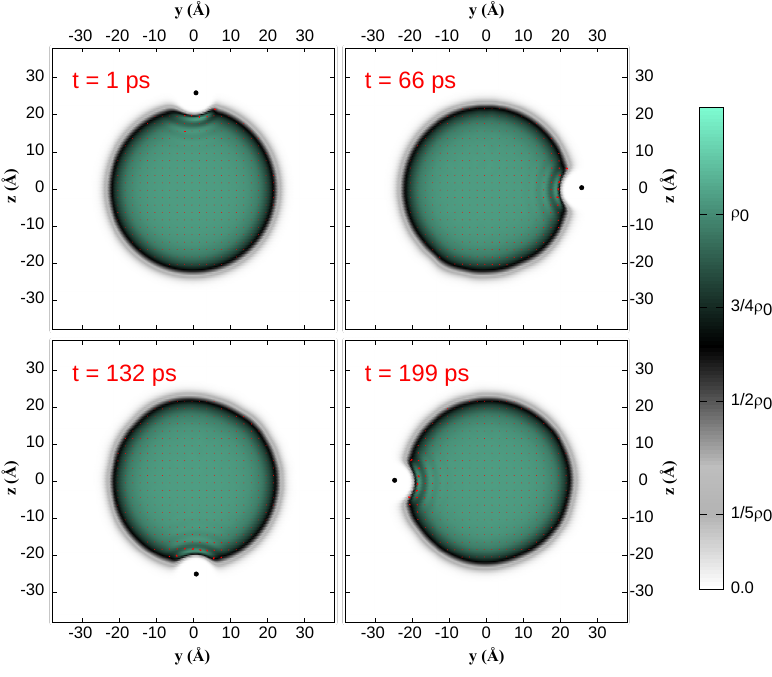}
\caption{
    \textbf{Dynamics of a Na atom moving on the surface of a $^4$He$_{1000}$ nanodroplet for an initial velocity of 70~$\mps$.}
    The Na atom is initially located in a dimple at the surface of a He nanodroplet, and its velocity is initialized parallel to the droplet surface (along the $y$ axis).
    Each panel shows a 2D cut in the $y$-$z$-plane.
    The different panels present snapshots at selected times after the ``kick'' imparted to the Na atom. Within the first few picoseconds, the Na atom collides with the dimple, thereby losing part of its initial kinetic energy.
    After this initial slowdown, the Na atom continues to move along the droplet surface at approximately constant speed.
    \label{fig:surf_motion_slow}
}
\end{figure}

Fig.~\ref{fig:surf_motion_slow} shows the evolution of the system for the case $\vna = 70~\mps$. The different panels show 2D cuts of the He density in the $y$-$z$-plane at selected times after the initial kick imparted to the Na atom. The velocity field of the He density is represented by small red arrows, which are more clearly visible when the figure is enlarged. After the initial slowdown due to the transfer of angular momentum from the Na atom to the dimple, we only see a small deviation from the spherical shape of the droplet surface as $\vna = 70~\mps$ is just above the threshold for ripplon excitation (see discussion below). Quantum vortices do not form due to insufficient transfer of angular momentum to the droplet for $\vna = 70~\mps$.

\begin{figure}
\centering
\includegraphics[width=0.8\linewidth]{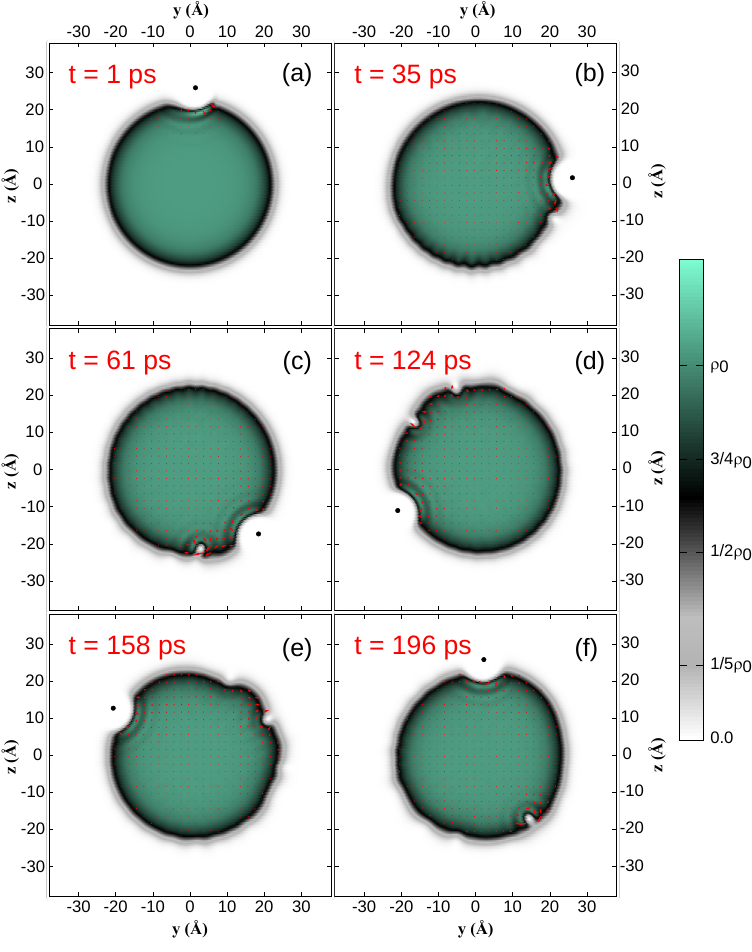}
\caption{
    \textbf{Dynamics of a Na atom moving on the surface of a $^4$He$_{1000}$ nanodroplet for an initial velocity of 150~$\mps$.}
    In contrast to Fig.~\ref{fig:surf_motion_slow}, the Na atom decelerates further after its initial slowdown caused by bouncing off the dimple. Concurrently, small voids with accelerated He flow around them (vortices) form on the droplet surface.
    \label{fig:surf_motion_fast}
}
\end{figure}

By contrast, when we increase $\vna$ to $150~\mps$, a markedly different droplet response is observed, as shown in Fig.~\ref{fig:surf_motion_fast}.
As in the previous case of $\vna = 70~\mps$ (Fig.~\ref{fig:surf_motion_slow}), the Na atom initially slows down due to its collision with the surrounding dimple. However, it continues to lose speed and, at the same time, a small dent develops at the droplet surface, as is best seen in panel (c). There, the He density features a void at the bottom of the droplet at $y= 2$ and $z= 20.9$~\AA. A fast circular flow of He in the vicinity of this void (see the red arrows) indicates that this void might be a vortex. At later times, panels (d)-(f), a second void forms in front of the first one, and the latter disappears again.

The slowdown of the Na atom for $\vna > 70~\mps$ can be explained by surface excitations that carry angular momentum.
By analogy with the critical velocity in bulk He-II, there is a critical angular velocity, $\omega_{cr}$, above which superfluidity breaks down and ripplons (quantized capillary waves) can be excited on the droplet surface
\cite{pitaevskiiSuperfluidEffectsRotating1990}. 
The critical angular frequency for ripplon excitation, $\omega_{cr}$, depends on the number of He atoms in the nanodroplet,
$N_\mathrm{He}$, and is given by~\cite{pitaevskiiSuperfluidEffectsRotating1990, tanyagCHAPTER8Experiments2017}
\begin{equation}
    \omega_{cr} = 667.69 \cdot N_\mathrm{He}^{-1/2}~[\mathrm{GHz}].
    \label{eq:cri_ripplons}
\end{equation}
The corresponding ripplon energy is~\cite{pitaevskiiSuperfluidEffectsRotating1990}
\begin{equation}
    E_r = 10.2 \cdot N_{\mathrm{He}}^{-1/2}~[\mathrm{K}].
\end{equation}
For $N_\mathrm{He} = 1000$, these expressions yield $\omega_{cr} = 21.1~\mathrm{GHz}$ and $E_r = 322.5~\mathrm{mK}$.

Even below the threshold for ripplon excitation, $\omega_{cr}$, the droplet can in principle acquire angular momentum in the form of quantum vortices.
However, for a vortex to form, the system needs to overcome an energy barrier. A vortex passing through the center of a spherical droplet containing $N_\mathrm{He}$ He atoms carries the angular momentum $\angmom = N_\mathrm{He}\,\hbar$~\cite{pitaevskiiSuperfluidEffectsRotating1990}.
The corresponding vortex energy~\cite{bauerVortexConfigurationsFreely1995, lehmannEnergeticsPossibleFormation2003} is
\begin{equation}
    E_v = 3.69 N^{1/3}_\mathrm{He} \left[ \ln \left( 4.44 N^{1/3}_\mathrm{He} \right)  - 1\right]~[\mathrm{K}],
\end{equation}
and the critical angular velocity for vortex nucleation, $\omega^v_{cr}$, is given by~\cite{pitaevskiiSuperfluidEffectsRotating1990}
\begin{equation}
    \omega^{v}_{cr} = 157 N_\mathrm{He}^{-2/3} \mathrm{ln}\left(N_\mathrm{He}\right)~[\mathrm{GHz}].
    \label{eq:cri_vortex}
\end{equation}
For $N_\mathrm{He} = 1000$, we obtain $\angmom = 1000\,\hbar$, $E_v = 103.07~\mathrm{K}$, and
$\omega^{v}_{cr} = 10.8~\mathrm{GHz}$.

For a Na atom with an initial velocity $\vna = 70~\mps$ and an initial radial position
$\rna = 25.65~\text{\AA}$ relative to the droplet center,
the angular velocity of the Na atom is
\begin{equation}
    \omega_\mathrm{Na} = \vna / \rna = 27~\mathrm{GHz},
\end{equation}
and the kinetic energy is
\begin{equation}
    E_{\mathrm{Na}} = \frac{1}{2} m_{\mathrm{Na}} v^2_{\mathrm{Na}} = 6.7~\mathrm{K}.
\end{equation}
The kinetic energy, $E_{\mathrm{Na}}$, exceeds the ripplon excitation energy, $E_r$.
The angular velocity, $\omega_{\mathrm{Na}}$, is also quite close to the ripplon threshold, $\omega_{\mathrm{cr}}$,
and higher than the threshold for vortex formation, $\omega^v_{\mathrm{cr}}$.
Thus, an initial Na speed $\vna > 70~\mps$ can excite ripplons and thereby slow the Na motion.
If sufficient angular momentum is transferred from the Na atom to the droplet, vortices can also form.

To assess whether the moving voids observed for $\vna=150~\mps$\ in Fig.~\ref{fig:surf_motion_fast}, panels (c) to (f), are indeed vortices, we analyze the He flow field around them.
In He-II, the circulation around a vortex,
i.\,e., the line integral of the fluid velocity along a closed loop that encircles it, is quantized
\cite{feynmanChapterIIApplication1955},
\begin{equation}
    \oint_c \mathbf{v} \cdot d \mathbf{\ell} = n  \frac{2 \pi \hbar} {m_\mathrm{He}} = n \kappa.
\end{equation}
Here, $n$ is the quantum number, $\kappa=2 \pi \hbar/m_\mathrm{He}$ is the quantum of circulation, and $m_\mathrm{He}$ is the mass of a $^4$He atom.
For the void located at the bottom of the He droplet in Fig.~\ref{fig:surf_motion_fast}(c), the circulation around the singularity equals $\kappa$ with $n=1$, which confirms that the void is a vortex.
A three-dimensional view of this curved surface vortex is shown in Fig.~\ref{sfig:surface_vortex}.

This finding may appear surprising at first glance, as for $\vna = 150~\mps$ the values $\omega_\mathrm{Na} = 58~\mathrm{GHz}$, $L_\mathrm{Na} = 139\hbar$, and $E_\mathrm{Na} = 31.12~\mathrm{K}$ are insufficient to excite a vortex passing through the center of the droplet. However, much lower values for the angular momentum and energy are needed to excite a vortex near the droplet surface. The reduction in angular momentum and energy needed for an off-center vortex
can be understood geometrically.
A vortex line passing through the droplet center generates circulation throughout the entire fluid volume surrounding the line. As the vortex core is displaced toward the boundary, the fraction of the fluid volume that participates in the circulation decreases, which in turn reduces the angular momentum carried by the vortex~\cite{bauerVortexConfigurationsFreely1995}.

The vortex line must satisfy the boundary conditions at the droplet surface and is therefore expected to intersect the interface normally.
As a consequence, only a vortex that passes through the center can remain straight, whereas an off-center vortex must bend in order to satisfy the surface boundary condition
~\cite{lehmannEnergeticsPossibleFormation2003,
bauerVortexConfigurationsFreely1995}.
Numerical calculations by Lehmann~\textit{et al.}
\cite{lehmannEnergeticsPossibleFormation2003} support this assumption.
By fitting their numerical results (Fig.~4 in \cite{lehmannEnergeticsPossibleFormation2003}), one obtains the empirical 
scaling relation
\begin{equation}
\angmom = \angmom^0 \left(1-r^2/R^2\right)^n,
\label{eq:angmom_scaling}
\end{equation}
with $n=2.4$.
In Eq.~(\ref{eq:angmom_scaling}), $\angmom^0$ is the angular momentum for a vortex passing through the droplet center,
$r$ is the vortex's distance from the droplet center, and $R$ is the droplet radius.
A more detailed derivation of the scaling law and a comparison with Ref.~\cite{lehmannEnergeticsPossibleFormation2003} are given in the SM and
Fig.~\ref{sfig:angmom_scaling}.
This result generalizes the formula for a freely rotating cylindrical drop [$\angmom\propto (1-r^2/R^2)$] established by Bauer et al.~\cite{bauerVortexConfigurationsFreely1995}.
Because Eq.~(\ref{eq:angmom_scaling}) was obtained for a spherical droplet hosting a curved vortex, it is not directly applicable to the present simulations, in which the droplet surface deviates from sphericity due to the presence of ripplons  (see Fig.~\ref{fig:surf_motion_fast}).
Nevertheless, within the idealized assumption of a spherical droplet
with a vortex line intersecting the surface normally,
a vortex at $r=20.9~\text{\AA}$ 
(corresponding to $r\simeq0.94R$ in Fig.~\ref{fig:surf_motion_fast}(c)) would carry only $\angmom\simeq5.7\hbar$.
Hence, an angular-momentum transfer of $\angmom>5.7\hbar$ from the orbiting Na atom to the He nanodroplet is sufficient to nucleate a vortex located near the droplet surface.

In Fig.~\ref{fig:surf_motion_fast}(c), the vortex is formed at a position $r = 0.94 R$. The only way for the droplet to host a stable vortex is to carry
enough angular momentum so that
the vortex can move inside the droplet to a distance $r < 0.76 R$
\cite{lehmannEnergeticsPossibleFormation2003}.
As this would require the transfer of $\angmom > 131 \hbar$
and $E_v > 27.1~\mathrm{K}$ to the droplet, the vortex remains unstable.
Indeed, the formation of a vortex is followed by a back-and-forth transition between vortices and ripplons;
see Fig.~\ref{fig:surf_motion_fast} and Movie~S2 in the SM.
In this dance of vortices and ripplons, the vortices periodically lose angular momentum by transfer of angular momentum to ripplons and back to vortices.
Fig.~\ref{fig:surf_motion_fast}(d) shows the initial vortex at
a radius $r = 22.02~\text{\AA}$ moving toward the droplet surface and transferring some of its $\angmom$ to ripplons.
In turn, a new void starts to form just in front of the initial vortex. Eventually, the initial vortex completely decays and the void becomes a vortex.
Fig.~\ref{fig:surf_motion_fast}(e) shows the newly formed vortex
at $r = 22.5~\text{\AA}$, just inside the droplet.
As this vortex gains more angular momentum, it moves deeper inside the droplet.
This can be seen in Fig.~\ref{fig:surf_motion_fast}(f)
where the vortex is located again at $r=20.9~\text{\AA}$ after $t = 196~\mathrm{ps}$.
Figure~\ref{sfig:surf_vel} in the SM compares the surface dynamics for several initial velocities, illustrating the transition from weak surface deformation to coupled ripplon--vortex dynamics.

\subsection*{Motion inside the droplet}

\begin{figure}
\centering
\includegraphics[width=0.7\linewidth]{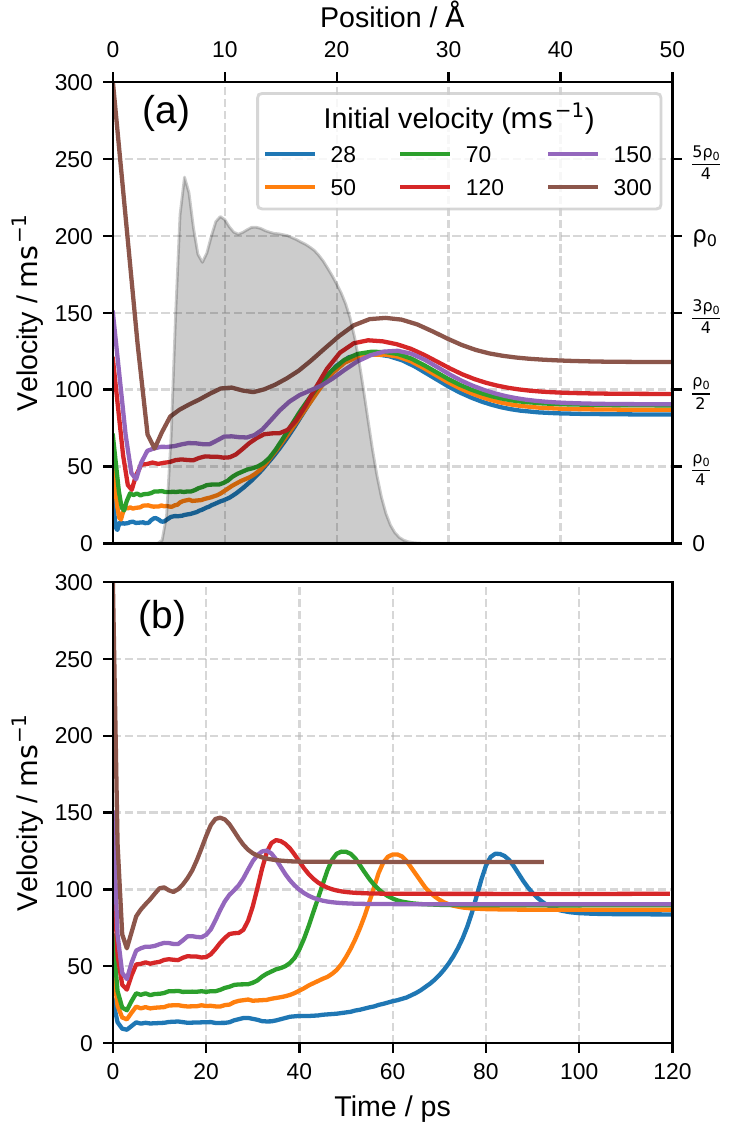}
\caption{
    \textbf{Motion of the Na atom through the interior of a $^4$He$_{1000}$ nanodroplet.}
    Differently colored lines show the evolution of the Na atom's velocity $\vna$
    for different initial velocities.
    Panel (a) shows $v_\mathrm{Na}$ as a function of $r_\mathrm{Na}$ with respect to the droplet center (top $x$-axis). The shaded area represents the He density profile for the static case (right $y$-axis). Panel (b) shows $v_\mathrm{Na}$ as a function of simulation time.
    \label{fig:bulk_vel}
}
\end{figure}

To compare our findings with the previously studied motion in
the bulk of He nanodroplets~\cite{brauerCriticalLandauVelocity2013,lealCaptureHeliophobicAtoms2014,vilaQuantumDynamicsPick2016,coppensHeadonCollisionsXe2017,coppensCaptureXeAr2017,garcia-alfonsoClusteringCollisionRelaxation2022,bonhommeauCollisionCesiumAtoms2024}, 
we carried out a set of simulations for the same impurity, Na, starting at the droplet center. 
The corresponding metastable configuration with the Na atom constrained to the center of the He nanodroplet was obtained by static DFT simulations with the addition of a constraint potential \cite{ancilottoDensityFunctionalTheory2017}.
It was then taken as the initial geometry for the dynamic simulation, and the velocity of the Na impurity $\vna$ was initialized along the $y$-direction.
As a result, at $t = 0$ the Na atom sits inside a bubble of radius $5.45~\text{\AA}$ at the droplet center, and it soon collides with the surrounding He density; see Fig.~\ref{sfig:static}(b) and the Materials and Methods for more details.

Fig.~\ref{fig:bulk_vel} shows the evolution of the velocity $\vna(t)$ of the Na atom for selected initial velocities, plotted against the distance of the Na atom from the droplet center $\rna$ in (a) and against the simulation time in (b). Each colored line shows the trajectory for a different initial velocity.
The shaded region depicts the density profile of the He droplet for the initial configuration.

All trajectories show an initial rapid slowdown of the Na atom within the first $\sim5$~ps as the Na bounces off the He bubble that surrounds it.
Subsequently, the He bubble and the Na atom are slowly accelerated toward the droplet surface due to the repulsive short-range interaction of the Na atom with the He droplet.
As the bubble gets to the droplet surface,
it is strongly accelerated and subsequently bursts,
thus ejecting the Na atom from the droplet.
In the last phase of the interaction of the Na atom
with the He droplet,
the Na atom is again slightly slowed down due to attractive
long-range interactions of the Na atom with the He droplet [see, e.g., Figs.~1 and 2 in ref.~\cite{lealCaptureHeliophobicAtoms2014}]. Subsequently, the Na atom continues to move away from the droplet with an almost constant speed of about $90~\mps$.
Only for an initial velocity $\vna = 300~\mps$
is the final Na velocity slightly higher than $90~\mps$, probably due to more violent Na-droplet interactions leading to
more ballistic motion of the Na atom.
In the early phase following the initial strong deceleration, when the Na atom moves inside the He droplet (see the velocity plateau after 5~ps in Fig.~\ref{fig:bulk_vel}(b)), its average velocity continuously increases from about $10$ to $100~\mps$ as
the initial velocity ranges from $28$ to $300~\mps$,
respectively. This is inconsistent with the existence of a critical velocity below which vanishing friction would be exerted,  as in bulk He-II.
Note, however, that the value of $v_L$ for the functional used in this work ($v_L=118~\mps$) is higher than the experimental value (see the Discussion section below).

\begin{figure}
\centering
\includegraphics[width=0.7\linewidth]{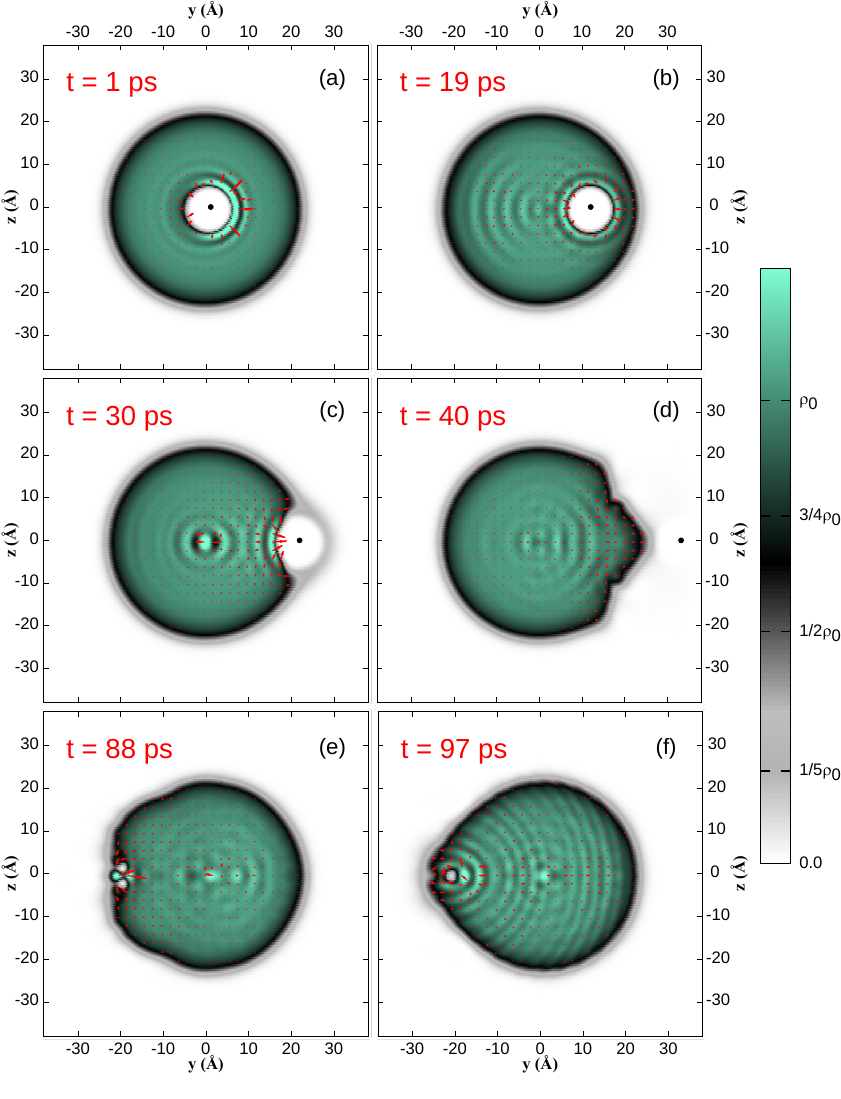}
\caption{
    \textbf{Dynamics of a Na atom moving through a $^4$He$_{1000}$ nanodroplet with
    an initial velocity of $150~\mps$.}
    The motion is initialized at the droplet center with an initial velocity in the $y$ direction.
    Panels (a)-(f) show 2D cuts of the He density in the $y$-$z$-plane at different simulation times. Within the first few picoseconds, the Na atom collides with the surface of the bubble that surrounds it, thereby losing most of its kinetic energy. Subsequently, the Na atom is accelerated toward the droplet surface and is ejected [panel (c)]. The collapse of the bubble creates two counter-propagating surface waves (d) that interfere with each other on the opposite side of the droplet (f).
    \label{fig:bulk_motion}
}
\end{figure}

The response of the He density as the Na impurity moves through
it is visualized in Fig.~\ref{fig:bulk_motion}.
It shows the motion of the Na atom inside the He droplet with an initial velocity
$\vna = 150~\mps$ as 2D cuts of the He density in
the $y$-$z$-plane at selected simulation times.
The velocity field of the He density is again shown as red arrows. Just after the start of the simulation, the Na atom bounces off the surface of the bubble surrounding it, thus losing most of its initial kinetic energy,
$E_\mathrm{Na} = 31.12~\mathrm{K}$, to the He droplet.
As the initial configuration is unstable, the bubble containing the Na atom accelerates toward the droplet surface,
driven by the repulsion of the Na from the surrounding
He, see panel (b). When the bubble crosses the He-vacuum interface, it collapses and the Na atom is ejected from the droplet, panel (c).
This induces two counter-propagating surface waves, panels (d) and (e),
which run along the upper and lower sides of the droplet and
eventually interfere with each other and excite bulk excitations in the droplet, panels (e) and (f).
In particular, a vortex ring is excited at $t = 88~\mathrm{ps}$.
A semi-transparent $x$-axis view of the vortex ring is shown in
Fig.~\ref{sfig:bulk_vortex}.
Additional interior trajectories are shown in Fig.~\ref{sfig:bulk_vel}.

\section*{Discussion}

In the case of the Na atom moving along the He nanodroplet surface, we identified a distinct critical velocity $v_c\approx70~\mps$ for undamped motion,
corresponding to an angular frequency of about 27~GHz for the He$_{1000}$ droplet considered here. Below this velocity, the Na atom is initially decelerated due to its interaction with the dimple it resides in, followed by essentially frictionless motion along the droplet surface. Above $v_c$, excess kinetic energy and angular momentum are transferred from the Na atom to the He droplet, resulting in the excitation of ripplons and, accordingly, in damping of the Na motion. At sufficiently high velocities, $\vna = 150~\mps$, vortices are nucleated near the droplet surface. These vortices are unstable and dynamically coupled to the ripplon excitations, with angular momentum being transferred
back and forth between the two types of excitation.

The critical velocity for surface motion we found is substantially lower than the Landau critical velocity determined previously for motion through bulk He-II using the same density functional as in this work (which omits both the backflow and the kinetic-energy correlation terms to save CPU time), $v_L = 118~\mps$, or $\omega_L = v_L / \rna = 46~\mathrm{GHz}$.
Omitting the backflow and kinetic-energy correlation terms increases the Landau critical velocity from $\sim58~\mps$ to $118~\mps$~\cite{ancilottoDensityFunctionalTheory2017}.
The difference between the critical velocities for surface and bulk motion highlights the important role of the finite size and free surface of He nanodroplets:
For an impurity moving along the surface,
ripplons and surface-bound vortices provide additional channels
for dissipation that are absent from the conventional picture of an impurity moving through bulk He-II. 
Our results therefore demonstrate that the concept of a critical velocity still applies for motion at the droplet surface, albeit with a smaller value than in bulk He II.
Therefore, a finite superfluid system cannot be characterized by the bulk Landau criterion alone.

The motion of impurity atoms inside He nanodroplets has previously been investigated theoretically mainly from the perspective of the pickup process. 
In a simulation of a Cs atom colliding with a He nanodroplet with a relatively high impact velocity of $200~\mps$, the atom pierced the droplet and subsequently lost nearly all its remaining kinetic energy by excitation of
a vortex ring in the droplet
\cite{lealCaptureHeliophobicAtoms2014,coppensHeadonCollisionsXe2017}.
Although it was slowed almost to a standstill in the droplet \cite{coppensHeadonCollisionsXe2017}, it was eventually expelled from the droplet with a final speed of $64.7~\mps$~\cite{lealCaptureHeliophobicAtoms2014}.
Coincidentally, this value is comparable to the experimental bulk He-II Landau critical velocity.
Similarly, our simulations also show that Na atoms are ejected from the droplets with nearly equal final speeds irrespective of their initial velocities.
Therefore, we interpret the final Cs velocity as resulting solely from the acceleration of
the Cs atom while traversing the region of inhomogeneous He density at the droplet surface during
the ejection process (the bursting of the Cs bubble).

For peripheral collisions of a Cs atom with a He nanodroplet at a lower velocity ($50~\mps$), it was reported that
the impinging Cs atom formed a dimple that rotated around the droplet,
carrying the Cs atom with it \cite{lealCaptureHeliophobicAtoms2014}.
Although the droplet appeared to rotate as a rigid body, the He flow remained irrotational;
the angular momentum was carried by capillary waves (superpositions of ripplons)
traveling on the droplet surface.
While the Cs velocity was significantly reduced in the process of attaching to the He nanodroplet
and orbiting around it for up to 1.3~ns, the limiting speed of the Cs atom and
the possible existence of a critical velocity were not investigated in that work.

Motivated by the experiment aimed at probing the Landau critical velocity in He nanodroplets \cite{brauerCriticalLandauVelocity2013},
the translational dynamics of an atom inside a He nanodroplet was previously investigated theoretically for the case of electronically excited silver, Ag$^*$, which, like alkali metals, is heliophobic~\cite{brauerCriticalLandauVelocity2013,mateoTranslationalDynamicsPhotoexcited2013}.
Similarly to our simulations of a Na atom initially located inside the droplet,
those simulations were initialized by placing the Ag$^*$ atom near the center of the nanodroplet (reflecting experimental conditions in which the heliophilic ground state atom was suddenly excited).
Although the Ag$^*$ had zero initial velocity, it was ejected from the droplet due to the repulsive Ag$^*$-He interaction.
A limiting speed of the ejected Ag$^*$ of $87~\mps$ was found, irrespective of the initial position of Ag$^*$ in the droplet.
This value is close to the Landau velocity of $v_L = 94~\mps$ for the He functional used in the same work (neglecting only the backflow term).
However, that limiting speed seemed to be unrelated to the speed of the Ag$^*$ atom in the He droplet interior~\cite{mateoTranslationalDynamicsPhotoexcited2013}.

In the present work, we conclusively determine, through systematic $^4$He-DFT simulations, that the velocity of an atom ejected from the nanodroplet is completely unrelated to a critical velocity for its motion in the bulk of the droplet. Instead, it is determined predominantly by the dynamics of the bursting bubble and
the inhomogeneous He density encountered during the ejection process.
Thus, our systematic study of the Na atom's motion starting inside the He droplet at various initial velocities and ending with its ejection confirms the earlier findings~\cite{mateoTranslationalDynamicsPhotoexcited2013,lealCaptureHeliophobicAtoms2014} that the final velocity of the ejected impurity is essentially constant and independent of its initial motion inside the droplet.
Our results challenge the interpretation of the experiment by Brauer et al.~\cite{brauerCriticalLandauVelocity2013}, which attributed the constant speed observed for impurity ejection to the existence of a critical velocity for the motion inside the nanodroplet.
Instead, we unambiguously show that the final speed is entirely governed by the bursting process of the bubble containing the impurity as it reaches the surface, and thus bears no relation to the motion inside the droplet.

Unfortunately, the current simulations cannot unambiguously validate or invalidate the concept of a Landau critical velocity for the motion in the bulk of He nanodroplets because a velocity of $\approx 118~\mps$ ($v_L$ for the functional used in this work) could not be reached without inducing massive perturbations of the droplet and near-ballistic motion of the Na atom. Therefore, we extended our simulations to larger droplets containing 2000 He atoms and initial Na velocities up to $v_\mathrm{Na}=800$~ms$^{-1}$;
see Fig.~\ref{sfig:bulk_vel_2000} in the SM for the velocity trajectories.
Figure~\ref{sfig:bulk_motion_800} shows the dynamics for the case of an initial speed of 800~ms$^{-1}$ featuring a reversal of the initial direction of Na motion after collision with the surrounding bubble, followed by its ejection from the droplet.
These simulations show that the Na atom always leaves the droplet with a constant velocity solely
determined by the dynamics of the bursting bubble at the droplet surface, thus confirming the picture of a constant velocity of the ejected atom irrespective of its velocity inside the droplet.

Several questions still remain open needing further systematic investigation: How does the surface critical velocity depend on droplet size, the mass and size of the impurity particle, and impurity–He interaction strength? Such studies could reveal whether the $\approx70~\mps$ threshold found here evolves toward a characteristic value in the macroscopic limit of a flat surface. It will also be important to establish the exact value of this critical velocity by repeating selected simulations using the more accurate Orsay-Trento functional that includes the backflow term~\cite{dalfovoStructuralDynamicalProperties1995}.  An intriguing question is by which mechanism are the ripplons and vortices interconnected and under which conditions surface vortices become stable and migrate into the droplet interior. More fundamentally, the relation between the surface critical velocity and the ripplon dispersion relation should be investigated, and distinct critical velocities associated with the excitation of ripplons, vortices, and rotons should be identified in view of a deeper theoretical understanding~\cite{skrbek2025critical}. Regarding the critical velocity for motion inside the droplet, more sophisticated simulations are needed to establish its existence conclusively.
In particular, larger droplets should be studied and previously neglected terms should be incorporated into the density functional, although both approaches are computationally expensive. To probe a larger range of velocities, the initial conditions could be prepared more carefully to realize a concerted motion of the impurity atom and the He bubble right from the start of the simulation and thus prevent the atom from bouncing into the He bubble and immediately slowing to low speeds.

Experimentally, the surface critical velocity could be investigated by a dedicated scattering experiment in which He nanodroplets collide with alkali atoms (e.\,g. Na, K) prepared in a beam with variable collision velocity, similar to pioneering crossed-beams experiments by Toennies and coworkers~\cite{lewerenzNewScatteringDeflection1993}. For relative velocities lower than the critical velocity, an alkali atom attaches to the droplet surface and orbits around the droplet without friction causing negligible energy dissipation in the droplet~\cite{lealCaptureHeliophobicAtoms2014}. For higher velocities, the surface motion of the attached atom is damped causing energy dissipation and evaporation of He atoms from the droplets. The resulting shrinkage of the droplets can be sensitively measured, e.\,g., as in Ref.~\cite{henning2019experimental}. Another approach would be to extend the time-resolved experiments mentioned in the Introduction, which probed the motion of He$^*$ atoms emerging to the droplet surface~\cite{sishodiaXUVFluorescenceProbe2026}, to alkali-doped He nanodroplets.

\section*{Materials and Methods}

The He density functional method in its static or time-dependent version,
$^4$He-(TD)DFT, is an important tool for modeling
the static and dynamic properties of superfluid He nanodroplets.
Since $^4$He-(TD)DFT works with the one-body density rather
than with the N-body wave function of He, it can be applied
to much larger systems than quantum Monte Carlo (QMC) or other ab initio quantum approaches.
This makes it possible to successfully connect with and
interpret a wide range of experimental results.
For a detailed description of the method, see the review articles
\cite{barrancoHeliumNanodropletsOverview2006,ancilottoDensityFunctionalTheory2017}.
In this work, we use the $^4$He-(TD)DFT method to obtain static equilibrium
geometries and to systematically study the motion of an impurity atom, Na, on and inside a He nanodroplet containing 1000 atoms (and inside a 2000-atom droplet in the SM).

In this approach, the total energy of the combined system of a Na
atom interacting with a He nanodroplet can be written as
\begin{equation}
    E_\mathrm{He}[ \Psi(\br) ]
    = \int d \br \frac{\hbar^2}{2m_\mathrm{He}} | \nabla \Psi (\br) |^2
    + \int d \br \mathcal{E}_c [\Psi(\br)]
    + \int d \br \mathcal{V}_\mathrm{He-Na} (| \br - \br_\mathrm{Na} |) \left| \Psi (\br)\right|^{2},
    \label{eq:total_energy}
\end{equation}
where the superfluid helium is treated as a single pseudo-wave function $\Psi(\br)$,
with the superfluid density given by $\rho(\br) = \left| \Psi(\br) \right|^2$,
and the normalization condition $\int d\br \left| \Psi(\br) \right|^2 = N_{\mathrm{He}}$.
Here, $N_{\mathrm{He}}$ is the number of He atoms in the droplet.

The first term of Eq.~\eqref{eq:total_energy} represents the kinetic energy
of an ideal Bose fluid with the same density as superfluid He.
The second term is the functional used in the calculation.
The functional is a modified version of the Orsay-Trento (OT) functional
\cite{dalfovoStructuralDynamicalProperties1995},
taking into account He--He correlations and finite-range interactions but omitting the computationally expensive backflow and kinetic-energy correlation terms; this version is often referred to as the `solid functional'
\cite{ancilottoFreezingHelium2005}.
The last term is added to introduce a Na impurity, treated as a classical particle, into the system.
It describes the interaction between He and the Na impurity, with
$\mathcal{V}_{\mathrm{He-Na}}$ being the Na--He pair potential reported
by Patil \textit{et al.}~\cite{patilAdiabaticPotentialsAlkali1991}.

The ground-state configuration of the He nanodroplet doped with the
Na impurity can be found by minimizing the total energy, Eq.~\eqref{eq:total_energy}, with respect to the pseudo-wave function $\Psi(\br)$.
To study the motion of the Na atom inside a He nanodroplet,
a constraint potential was added to Eq.~\eqref{eq:total_energy}
 in order to fix the distance between the Na atom and the center of mass (COM) of the nanodroplet \cite{ancilottoDensityFunctionalTheory2017}.
The Euler-Lagrange equation for the ground-state configuration of the system was solved numerically on a Cartesian grid of
$192\times192\times192$ points with a spacing of $0.4~\text{\AA}$
using the imaginary-time method
as implemented in the He-DFT BCN-TLS package~\cite{pi4HeDFTBCNTLSComputer2024}.

Fig.~\ref{sfig:static}(a) shows the static equilibrium configuration
for the He nanodroplet doped with a Na atom.
A Na atom resides in a dimple on the droplet surface $25.65~\text{\AA}$
away from the droplet center.
The Na solvation energy was found to be $-8.92~\mathrm{K}$, which agrees
well with the previously reported value~\cite{garcia-alfonsoTimeresolvedSolvationAlkali2024}.
The radius of the dimple in which the Na atom is located is $5.65~\text{\AA}$,
giving the dimple an effective mass of $m_d = 5.44~\text{amu}$;
see the SM.
Fig.~\ref{sfig:static}(b) shows the static metastable configuration
of the system with the Na atom fixed at the COM of the nanodroplet.
The Na solvation energy was found to be $70.46~\mathrm{K}$ (note the positive sign).
Due to the repulsive interaction between Na and He at short distances,
a bubble forms around the Na atom with a radius of $5.45~\text{\AA}$
and an effective mass of $m_{bub} = 59.2~\mathrm{amu}$;
see the SM.

The dynamic simulation was initialized from
the static equilibrium configuration 
by giving the Na impurity an initial ``kick'',
i.e., by setting its velocity component tangential
to the nanodroplet surface, $v_\mathrm{Na}$, to a finite value.
In the dynamics simulation,
the total energy of the system, Eq.~\ref{eq:total_energy},
is modified to include the kinetic energy of the Na impurity,
\begin{equation}
    E_{\mathrm{Na}} [\br_\mathrm{Na}, p_\mathrm{Na}] = \frac{p_\mathrm{Na}^2}{2m_\mathrm{Na}}.
    \label{eq:Na_ke}
\end{equation}
Here, $\br_\mathrm{Na}$ is the position of the Na atom in or on the He nanodroplet,
$p_\mathrm{Na}$ is the linear momentum and $m_\mathrm{Na}$ is the mass of the Na atom.

The Na impurity is treated as a classical particle
and the coupled equations governing the time evolution of the complex He wave function
and the Na atom's position are solved using
the $^4$He-TDDFT BCN-TLS package~\cite{pi4HeDFTBCNTLSComputer2024}
on the same Cartesian grid as the static simulation.
The $^4$He-TDDFT BCN-TLS package was ported from the Intel platform
to the more general GNU platform and its parallelization there was made more efficient.
This allowed us to perform systematic $^4$He-TDDFT simulations for a set of initial conditions.
A time step of 1~fs was found to be appropriate to maintain simulation stability. Additionally, an absorbing potential was applied at the boundaries of the simulation box to avoid spurious reflections of the helium wave function that could interfere with the correct dynamics.

\clearpage %

\section*{Acknowledgements}

K.S. thanks the ELI Beamlines Facility, a European user facility operated by the Extreme Light Infrastructure ERIC, for providing computational resources.
S.K. and K.S. thank the Department of Science and Technology as well as the Ministry of Education, Govt. of India, for partial financial support through research grants, and the Office of Global Engagement, IIT Madras, for support.
K.S. and M.M. thank F. Ancilotto and A. F. Vilesov for stimulating discussions. 

\section*{Author contributions}
K.S. performed the simulations and analyzed the data.
E.G.-A.H. and N.H. provided the code for the simulations, assisted in getting started and helped interpret the results.
S.K. contributed research resources, support and in manuscript preparation.
K.S. and M.M. wrote the draft of the manuscript. All authors read, commented on, and approved the manuscript.
\paragraph*{Funding:}
The research leading to this result has been supported by the COST Action CA21101 ``Confined Molecular Systems: From a New Generation of Materials to the Stars (COSY)''.
\paragraph*{Competing interests:}
The authors declare no competing interests.
\paragraph*{Data, code and materials availability:}
The datasets used and analyzed in the current study are available from the corresponding author upon request.

\subsection*{Supplementary materials}
Supplementary Text\\
Figs. S1 to S8\\
Movies S1 to S3

\newpage

\renewcommand{\thefigure}{S\arabic{figure}}
\renewcommand{\thetable}{S\arabic{table}}
\renewcommand{\theequation}{S\arabic{equation}}
\renewcommand{\thepage}{S\arabic{page}}
\setcounter{figure}{0}
\setcounter{table}{0}
\setcounter{equation}{0}
\setcounter{page}{1} %

\begin{center}
\section*{Supplementary Materials for\\ \scititle}
	Keshav~Sishodia$^{\ast}$,\
	Ernesto~Garc{\'i}a-Alfonso,\
	Nadine~Halberstadt,\\
    Sivarama~Krishnan,
    Marcel~Mudrich$^{\ast \ast}$\\
	\small$^{\ast}$Corresponding author. Email: keshav.sishodia@eli-laser.eu\\
    \small$^{\ast\ast}$Corresponding author. Email: mudrich@uni-kassel.de
\end{center}

\subsubsection*{This PDF file includes:}
Supplementary Text\\
Figs. S1 to S8\\
Captions for Movies S1 to S3

\newpage

\subsection*{Supplementary Text}

\subsubsection*{Static configurations}
\label{ssec:static_conf}

\begin{figure}[ht]
\centering
\includegraphics[width=0.8\linewidth]{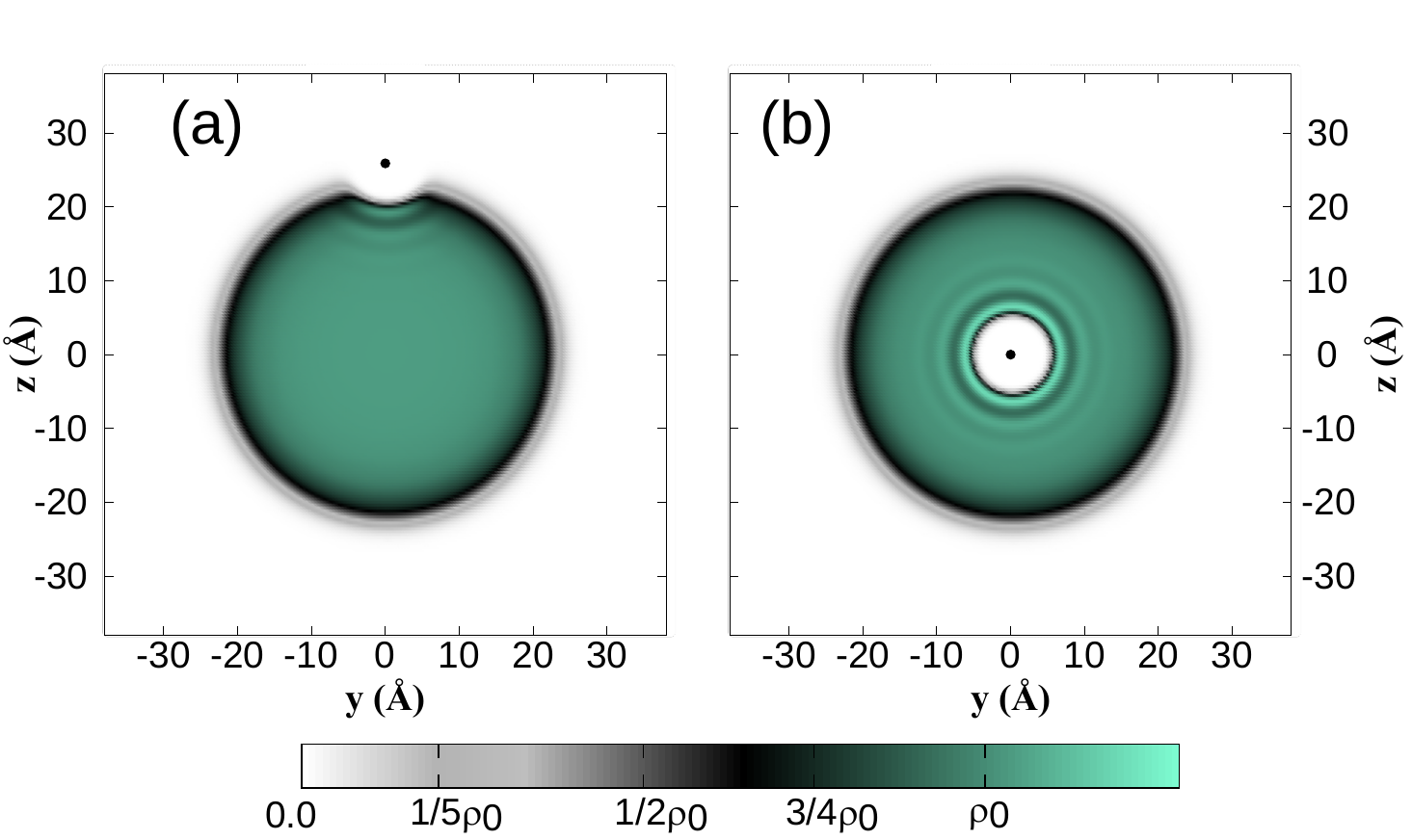}
\caption{
    \textbf{Static equilibrium configurations of a Na doped $^4$He$_{1000}$ nanodroplet.}
    Shown are 2D cuts through the He density distribution in the $y$-$z$-plane. (a) A Na atom initially placed on the surface of the nanodroplet resides in a shallow dimple $25.65~\text{\AA}$ from the droplet's center of mass. (b) A Na atom constrained to the droplet center by a constraint potential.
    \label{sfig:static}
}
\end{figure}

Figure~\ref{sfig:static} shows the static equilibrium configurations used as starting points for the simulations of Na motion on and inside the $^4$He$_{1000}$ nanodroplet, which has a sharp-density radius of $22.2$~\AA.
Figure~\ref{sfig:static}(a) shows the equilibrium configuration at the surface, where the Na atom resides $25.65$~\AA~from the droplet center in a dimple of radius $5.65~\text{\AA}$.
The corresponding Na solvation energy is $-8.92~\mathrm{K}$.
The interior configuration [Fig.~\ref{sfig:static}(b)] is a metastable equilibrium in which a void bubble of radius $5.45~\text{\AA}$ surrounds the Na atom at the droplet center.
The solvation energy of this configuration is $70.46~\mathrm{K}$.

\subsubsection*{Scaling law for the angular momentum of a He nanodroplet containing a vortex}
\label{ssec:scaling_law}

\begin{figure}
\centering
\includegraphics[width=0.8\linewidth]{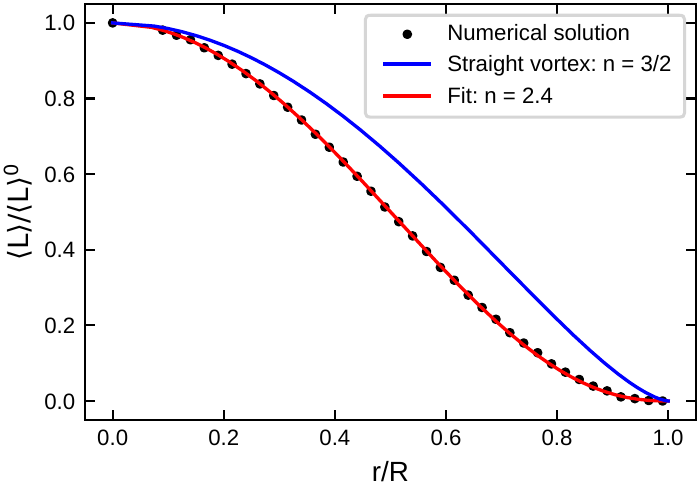}
\caption{
    \textbf{
    Scaling of the angular momentum $\langle L \rangle$ carried by a spherical droplet hosting an off-center vortex at a distance $r$ from the droplet center.}
    Here, $R$ denotes the droplet radius and $\langle L \rangle^0$ denotes the angular momentum of the droplet with a vortex passing through its center.
    The blue line shows the scaling law assuming that the off-center vortex is a straight line. The black dots are numerical data extracted from Ref.~\cite{lehmannEnergeticsPossibleFormation2003} and the red line shows the fit to these data.
    \label{sfig:angmom_scaling}
}
\end{figure}

As an extension of the derivation by Bauer et al.~\cite{bauerVortexConfigurationsFreely1995} for a cylindrical droplet, consider a spherical droplet of radius $R$ hosting a vortex at a distance $r_0$ from the center. The angular momentum of the system can be written as
\begin{equation}
    L = \int d^3 \br \rho(\br) (\br \times \bv),
    \label{seq:angmom_1}
\end{equation}
where $\rho(\br)$ is the He-II density and $\mathbf{v}$ is the He flow velocity.
To simplify the derivation, we make the following assumptions:
\begin{itemize}
    \item The $^4$He density, $\rho$, is constant throughout the He nanodroplet, i.e., it is independent of its distance $r$ from the center of the droplet.
    \item A vortex forms a straight line across the superfluid.
    \item $L$ is parallel to the $z$-axis. Therefore, we can replace $L$ with $L_z$.
\end{itemize}
Then, Eq.~\eqref{seq:angmom_1} can be rewritten as
\begin{equation}
    L_z = \rho \int d^3 \br~\hat{z} \cdot (\br \times \bv).
    \label{seq:angmom_approx}
\end{equation}
The scalar triple product $\hat{z}\cdot(\br \times \bv)$ can be written as $r\,v_{\phi}$,
where $v_{\phi}$ is the tangential component of the velocity vector.
Using cylindrical coordinates, this takes the form
\begin{equation}
    L_z = \rho \int_{-\sqrt{R^2 - r^2}}^{\sqrt{R^2 - r^2}} dz \int_0^R r dr \int_0^{2\pi} v_{\phi} r d\phi.
    \label{seq:angmom_polar}
\end{equation}
Here, $z$ is integrated along the vortex line.
The angular integral follows a closed circuit around the vortex.
For $r > r_0$, the circuit encloses the vortex and circulation is quantized
\cite{feynmanChapterIIApplication1955}.
For $r < r_0$, the circuit does not enclose the vortex and the angular integral becomes
zero.
\begin{equation}
    \int_0^{2 \pi} v_{\phi} r d \phi = \oint \bv \cdot d \mathbf{l} = 2\pi \frac{\hbar}{m}~\text{if}~r > r_0,~\text{otherwise}~ 0.
    \label{seq:circulation}
\end{equation}
Here, $\kappa = 2\pi \hbar/m$ is the quantum of circulation.
Substituting Eq.~\eqref{seq:circulation} into Eq.~\eqref{seq:angmom_polar}, we obtain
\begin{equation}
    L_z = 2 \pi \rho \frac{\hbar}{m} \int_{r_0}^R r dr \int_{-\sqrt{R^2 - r^2}}^{\sqrt{R^2 - r^2}} dz,
    \label{seq:angmom_final_integral}
\end{equation}
which yields the scaling law
\begin{equation}
    L_z = L_0 \left( 1 - \frac{r_0^2}{R^2}\right)^{3/2}.
    \label{seq:scaling_law}
\end{equation}

The scaling law in Eq.~\eqref{seq:scaling_law} is valid only when the 
off-center vortex line stays straight. As explained in the main manuscript,
this configuration is not physical because the vortex line must intersect the
droplet surface normally to satisfy the surface boundary condition. Consequently,
Eq.~\eqref{seq:scaling_law} should be regarded as an upper bound on the angular momentum. More accurate calculations by Lehmann~\textit{et.~al.}
indicate a different scaling behavior, in which the exponent in Eq.~\eqref{seq:scaling_law}
increases to 2.4; see Fig.~\ref{sfig:angmom_scaling}.

\subsubsection*{Effective mass of the dimple (Na at the droplet surface)}
\label{ssec:dimple_mass}

In the equilibrium configuration, the Na atom is bound within a dimple on the droplet surface, as shown in Fig.~\ref{sfig:static}~(a). By defining the dimple radius via the isodensity surface at which the He density is reduced to one half of its bulk value, we obtain a dimple radius of $r_d = 5.65~\text{\AA}$. To calculate the amount of He displaced by the dimple, we model the dimple as the intersection of two spheres: (i) the He droplet with radius $r_D = 22.2~\text{\AA}$ and (ii) a sphere with the radius of the dimple, $r_d$. The intersection volume is then given by
\begin{equation}
    V = \pi \left(r_D + r_d - d\right)^2
    \left(
    \frac{d^2 + 2d\,r_d - 3r_d^2 + 2d\,r_D + 6r_d\,r_D - 3r_D^2}{12d}
    \right),
    \label{seq:volume_intersection}
\end{equation}
where $d$ denotes the distance between the centers of the two spheres. Using this expression, the dimple volume is calculated as $V = 62.5~\text{\AA}^3$. Assuming the bulk of the He droplet has the density of superfluid He, the number of He atoms contained in this volume is
$N_{\mathrm{He}} = 1.36$, which corresponds to an effective dimple mass of
$m_d = 5.44\,\mathrm{amu}$.

This effective mass plays a key role in determining the early-time dynamics of Na motion on the droplet surface. At the beginning of the dynamical simulations,
the Na atom is assigned an initial velocity $v_{\mathrm{Na};\,i}$. The Na atom first interacts with the preformed dimple, and subsequently the Na atom and the dimple move collectively along the droplet surface with a common velocity. If the interaction between the Na atom and the dimple is approximated by an elastic collision
and we assume that, after the collision, the Na atom and the dimple travel
together with a final velocity $v_{\mathrm{Na};\,f}$,
conservation of kinetic energy yields
\begin{equation}
    v_{\mathrm{Na};\,f}
    = \sqrt{\frac{m_{\mathrm{Na}}}{m_{\mathrm{Na}} + m_d}}\,v_{\mathrm{Na};\,i}
    \simeq 0.9\,v_{\mathrm{Na};\,i}.
    \label{seq:vel_ratio_surface}
\end{equation}
This theoretical value is consistent with the velocity reduction observed in the simulations after the initial collision,
see Fig.~\ref{fig:surf_comb} in the main text. Hence, the Na--dimple collision provides a quantitative explanation for
the dominant part of the initial slowing down of the Na atom in its motion along the droplet surface.

\subsubsection*{Effective mass of the bubble (Na at the droplet center)}
\label{ssec:bubble_mass}

When the Na impurity is constrained at the center of the droplet, it generates a void around it, hereafter referred to as a bubble, due to the short-range repulsive interaction with helium. Using the same cutoff density as in the previous case, we obtain a bubble radius of $r_{bub} = 5.45~\text{\AA}$. The corresponding number of displaced He atoms is $N_\mathrm{He} = 14.8$, yielding an effective bubble mass of
$m_{bub} = 59.2~\text{amu}$.

Assuming once again an elastic collision between the Na atom and the surrounding bubble, the final velocity of the Na atom is given by
\begin{equation}
    v_{\mathrm{Na};\,f} = \sqrt{\frac{m_\mathrm{Na}}{m_\mathrm{Na} + m_{bub}}}~v_{\mathrm{Na};\,i} \simeq 0.53~v_{\mathrm{Na};\,i}.
    \label{seq:vel_ratio_bulk}
\end{equation}
However, inspection of Fig.~\ref{fig:bulk_vel} reveals that the measured final velocity is significantly lower than $53\,\%$ of the initial Na velocity. Therefore, the initial interaction between the impurity and the bubble cannot be accurately described as a purely elastic collision: A finite fraction of the kinetic energy of the impurity must be transferred to the He droplet.

\subsubsection*{Dynamics of a Na atom on the surface of a He nanodroplet}
\label{ssec:dynamics_surface}

\begin{figure}
\centering
\includegraphics[width=1.0\linewidth]{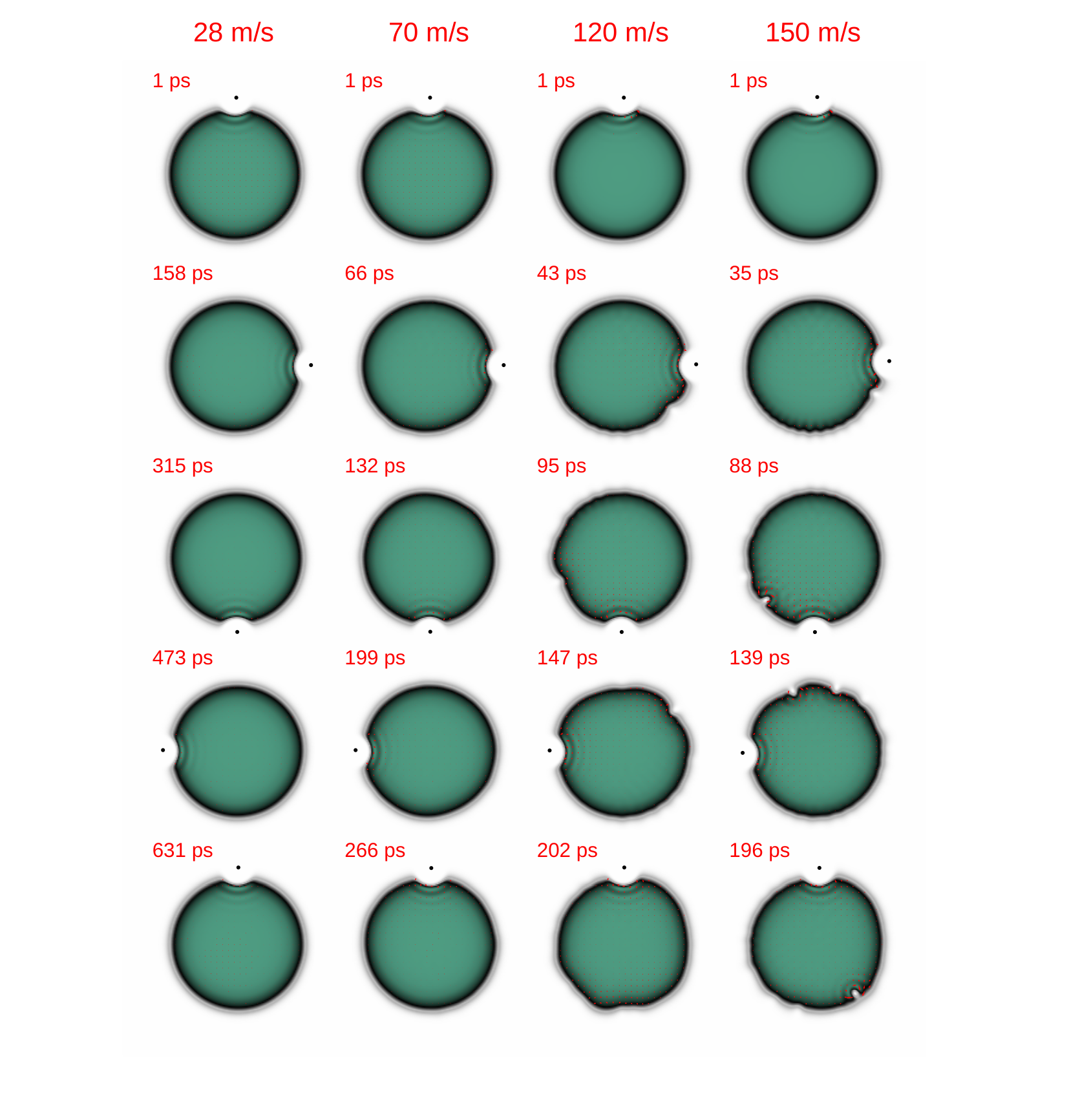}
\caption{
    \textbf{
    Dynamics of a Na atom on the surface of a $^4$He$_{1000}$ nanodroplet for different initial velocities.
    }
    The panels show a 2D cut of the He density in the $y$-$z$ plane.
    \label{sfig:surf_vel}
}
\end{figure}

Figure~\ref{sfig:surf_vel} illustrates the dynamics of the Na impurity on the He droplet surface for several initial velocities. A two-dimensional cross section of the He density is shown in the $y$-$z$ plane. For $v_{Na} \leq 70~\text{m}\,\text{s}^{-1}$, the initial deceleration of the Na atom primarily excites surface ripplons, while the overall droplet shape remains nearly spherical.
For $v_\mathrm{Na} > 70~\text{m}\,\text{s}^{-1}$,
the additional momentum transfer induces coupled dynamics involving ripplons and vortices, the latter seen as moving voids on the droplet surface.

\subsubsection*{Dynamics of a Na atom inside a He nanodroplet}
\label{ssec:dynamics_inside}

\begin{figure}
\centering
\includegraphics[width=1.0\linewidth]{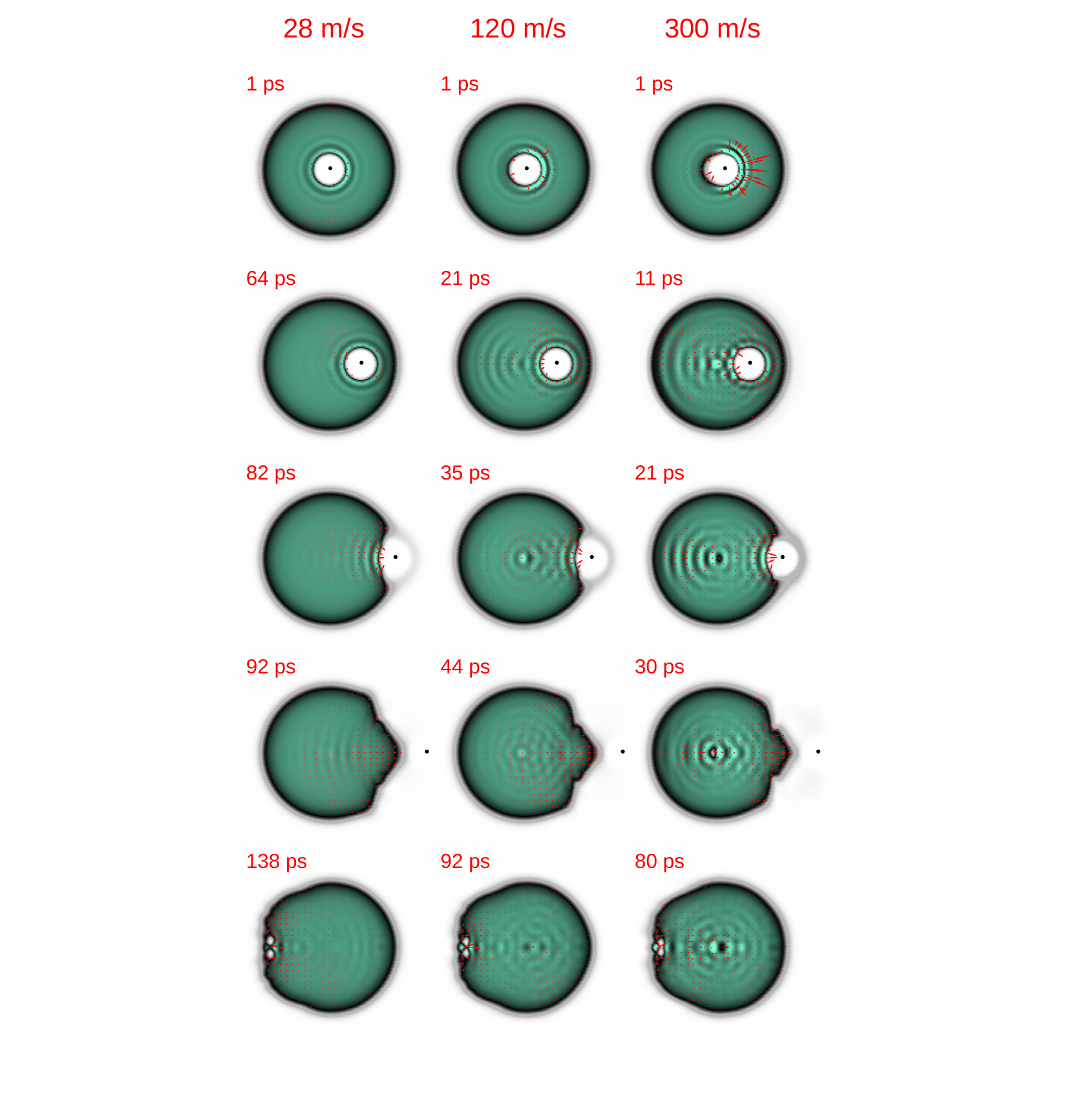}
\caption{
    \textbf{
    Dynamics of a Na atom moving inside a $^4$He$_{1000}$ droplet for different initial velocities.
    }
    The panels show 2D cuts of the He density in the $y$-$z$ plane.
    \label{sfig:bulk_vel}
}
\end{figure}

Figure~\ref{sfig:bulk_vel} presents the dynamics of a Na impurity inside a He nanodroplet for several initial impurity velocities. A two-dimensional cross section of the He density is displayed in the $y$-$z$ plane, with red arrows indicating the superfluid current within the He droplet. 
The qualitative character of the Na motion inside the droplet is essentially the same for all the velocities considered in this work.
The impurity proceeds through the same sequence of dynamical stages. Initially, the Na atom collides with the bubble that encloses it,
thereby transferring a substantial fraction of its initial kinetic energy to the bubble and the surrounding He density. Subsequently, the Na atom accelerates toward the droplet surface, driven by the repulsive interaction between Na and its bubble and the nearby He density. Once the bubble reaches and crosses the He-vacuum interface, it collapses, and the Na atom is expelled from the droplet. This bursting process generates
two surface waves that propagate in opposite directions, traveling along the upper and lower surfaces of the droplet and ultimately interfering with one another,
thereby exciting bulk modes within the droplet.

\subsubsection*{Vortex configuration in a superfluid He nanodroplet}
\label{ssec:vortex_conf}

\begin{figure}
\centering
\includegraphics[width=0.8\linewidth]{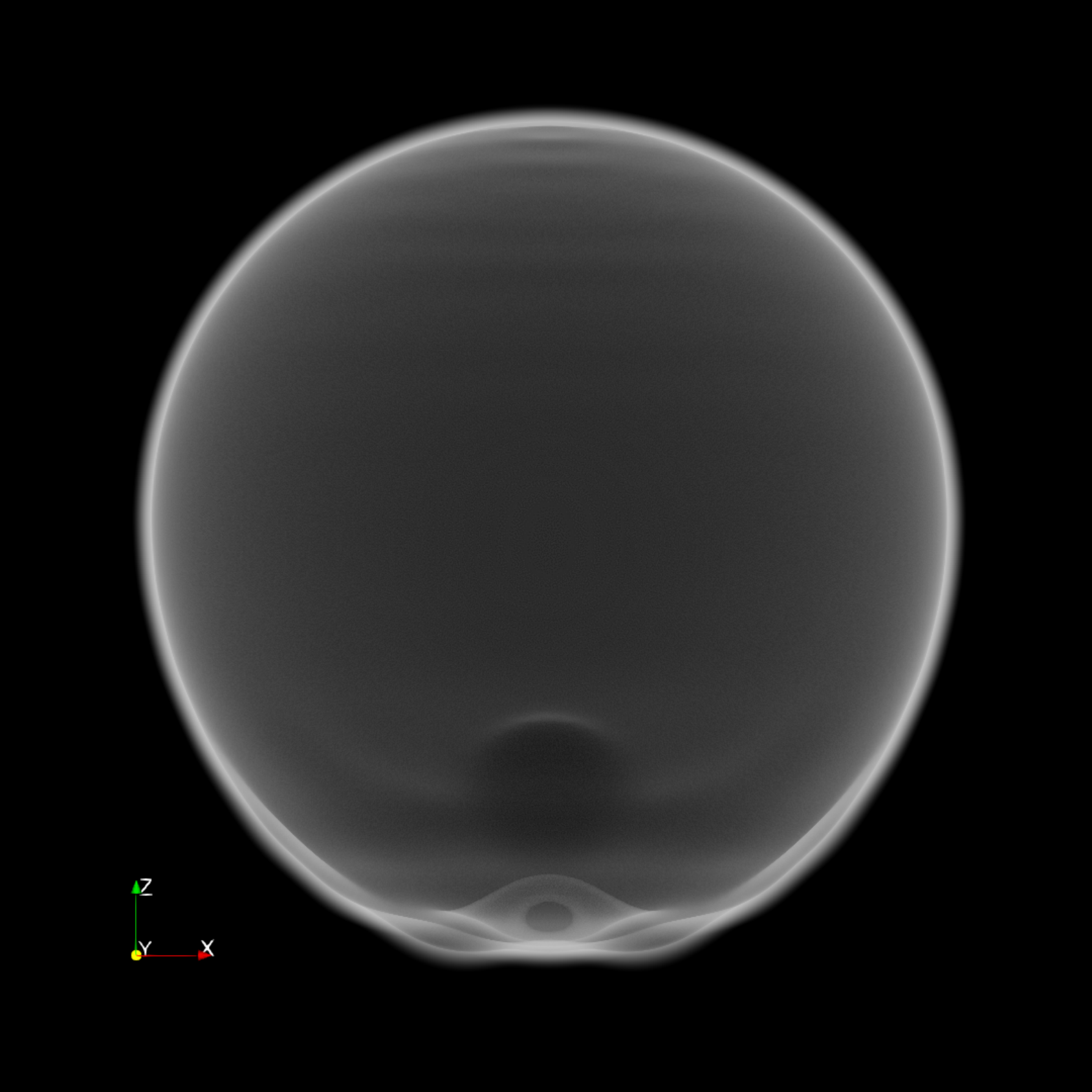}
\caption{
    \textbf{3D He droplet density snapshot at 61 ps for the dynamic simulation of a Na atom moving on the droplet surface with an initial velocity of 150~$\mps$.}
    In contrast to Figs.~\ref{fig:surf_motion_slow}~and~\ref{fig:surf_motion_fast}, here the He density is represented by its projection onto the $x$-$z$-plane. It is rendered in semi-transparent gray scale to visualize the surface dimple (darker round area in the bottom half) and a surface vortex (curved structure with a round darker spot at the bottom of the He density distribution).
    \label{sfig:surface_vortex}
}
\end{figure}

The vortex structure formed at $t = 61~\mathrm{ps}$ shown in
Fig.~\ref{fig:surf_motion_fast}(c) can be visualized
as a 3D rendered view along the $y$-axis, see Fig.~\ref{sfig:surface_vortex}.
Note that the view in Fig.~\ref{sfig:surface_vortex} is in
the $x$-$z$-plane, which is orthogonal to the $x$-$y$-plane used in
Figs.~\ref{fig:surf_motion_slow} and \ref{fig:surf_motion_fast}.
The dark round area in the middle of the lower half of the figure represents
the dimple that hosts the Na atom.
Just below it, one can see a curved hollow structure
that starts and ends at the droplet surface.

\begin{figure}
\centering
\includegraphics[width=0.8\linewidth]{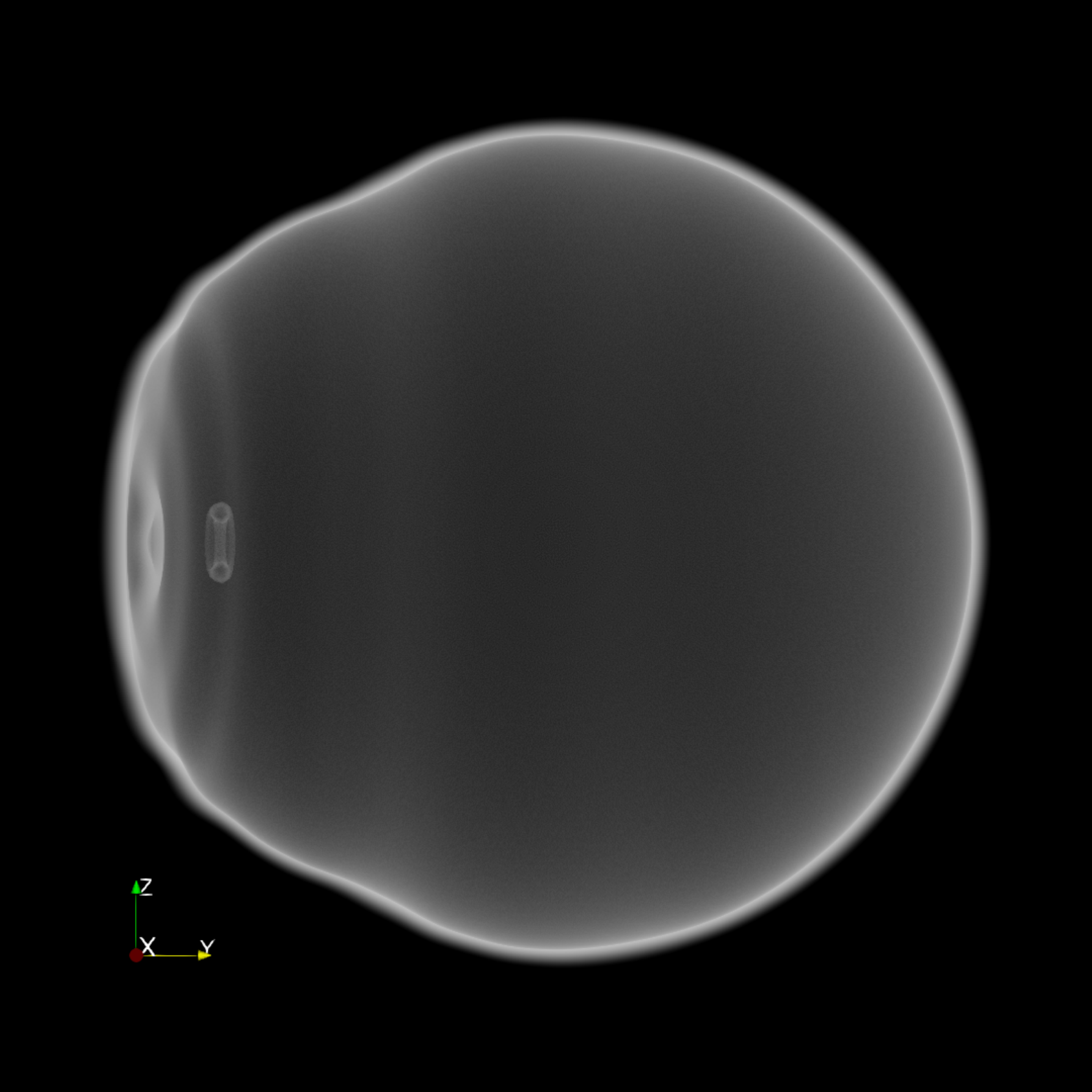}
\caption{
    \textbf{3D He droplet density snapshot at 88 ps for the dynamic simulation of a Na atom moving through the droplet with an initial velocity of 150~$\mps$.}
    The projection onto the $y$-$z$-plane is shown.
    The He density is displayed semi-transparently to visualize the vortex ring inside the droplet.
    \label{sfig:bulk_vortex}
}
\end{figure}

The vortex structure formed at $t = 88~\mathrm{ps}$ shown in
Fig.~\ref{fig:bulk_motion} can be visualized as a 3D rendered view along
the $x$-axis, see Fig.~\ref{sfig:bulk_vortex}.
In this semi-transparent gray-scale representation of the He density projection onto the $y$-$z$-plane, a vortex ring is seen as a small hollow donut near the left-hand side of the droplet just underneath the droplet surface. The vortex ring eventually collapses and the droplet relaxes back to its ground state.

\subsubsection*{Motion inside a He$_{2000}$ droplet}
\label{ssec:bulk_2000}

To assess whether the final velocity of the Na atom depends solely on the bubble–rupture mechanism, we performed simulations for a larger He droplet containing 2000 He atoms and initial Na velocities
up to 800~ms$^{-1}$.

\begin{figure}
    \centering
    \includegraphics[width=\linewidth]{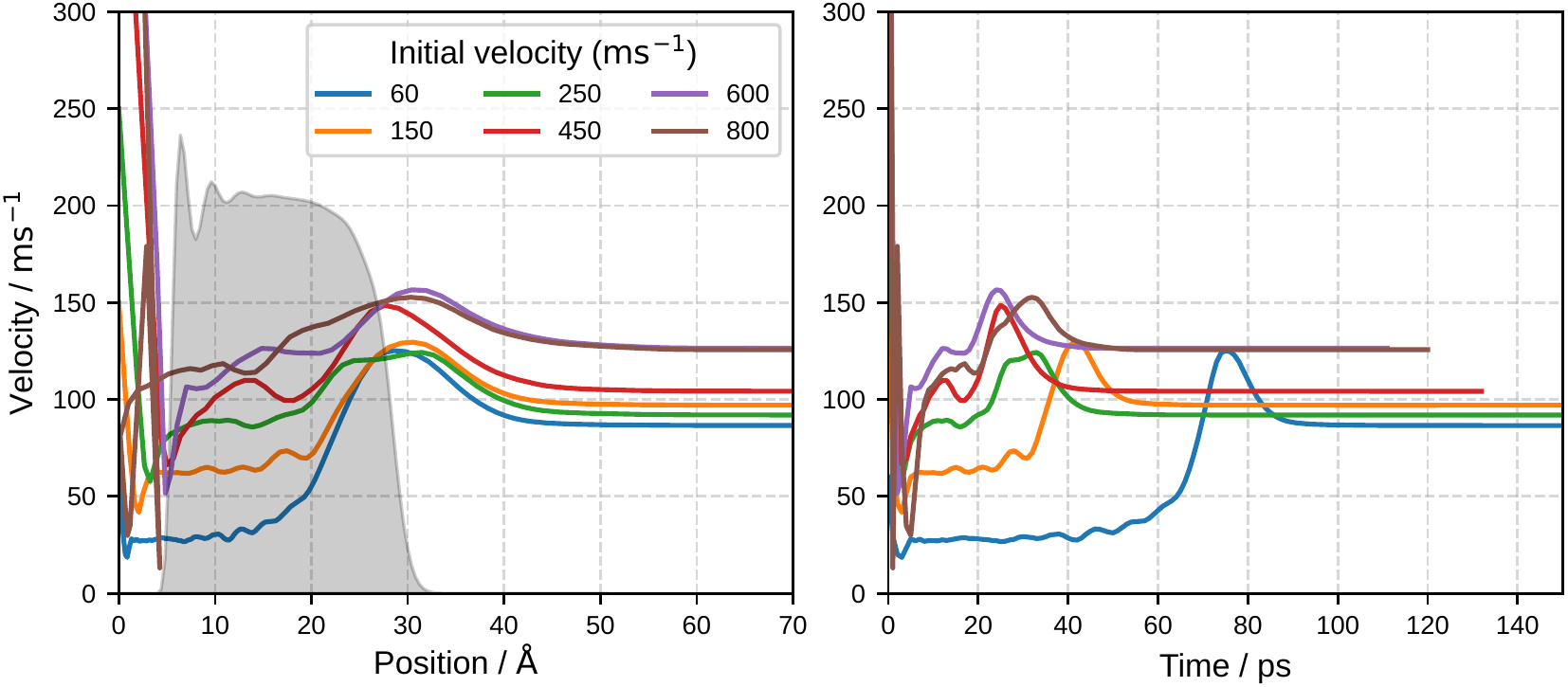}
    \caption{
        \textbf{Motion of the Na atom moving through the interior of a $^4$He$_{2000}$ nanodroplet.}
        Differently colored lines show the evolution of the Na velocity $\vna$ for
        different initial velocities.
        (a) $v_{\mathrm{Na}}$ as a function of the Na position $r_{\mathrm{Na}}$ with respect to the droplet center. The shaded area represents a one-dimensional cut of the He density for the static configuration. (b) $v_{\mathrm{Na}}$ as a function of time. The $y$-axis is restricted to 300~ms$^{-1}$ for clarity of visualization. For the initial speed of 800~ms$^{-1}$ (brown curve), the atom reverses direction after colliding with the surrounding bubble at approximately 1~ps; the trajectory is shown in Fig.~\ref{sfig:bulk_motion_800}.
    }
    \label{sfig:bulk_vel_2000}
\end{figure}

Fig.~\ref{sfig:bulk_vel_2000} displays $v_{\mathrm{Na}}$ as a function of
$r_{\mathrm{Na}}$ and of the time elapsed from the start of the simulation.
Each color corresponds to the velocity evolution for a different initial
value of the Na velocity. The shaded region in Fig.~\ref{sfig:bulk_vel_2000}~(a) indicates a one-dimensional cut of the
initial He density profile. The impurity dynamics are qualitatively similar to those observed for the He$_{1000}$ droplet. In particular, the Na atom undergoes the same three characteristic stages: (i) early loss of kinetic energy ($t\lesssim 5$~ps), (ii) slow acceleration toward the droplet surface, and (iii) rupture of the surrounding bubble,
followed by ejection from the droplet with an approximately constant asymptotic velocity, common to all initial velocities.

\begin{figure}
    \centering
    \includegraphics[width=\linewidth]{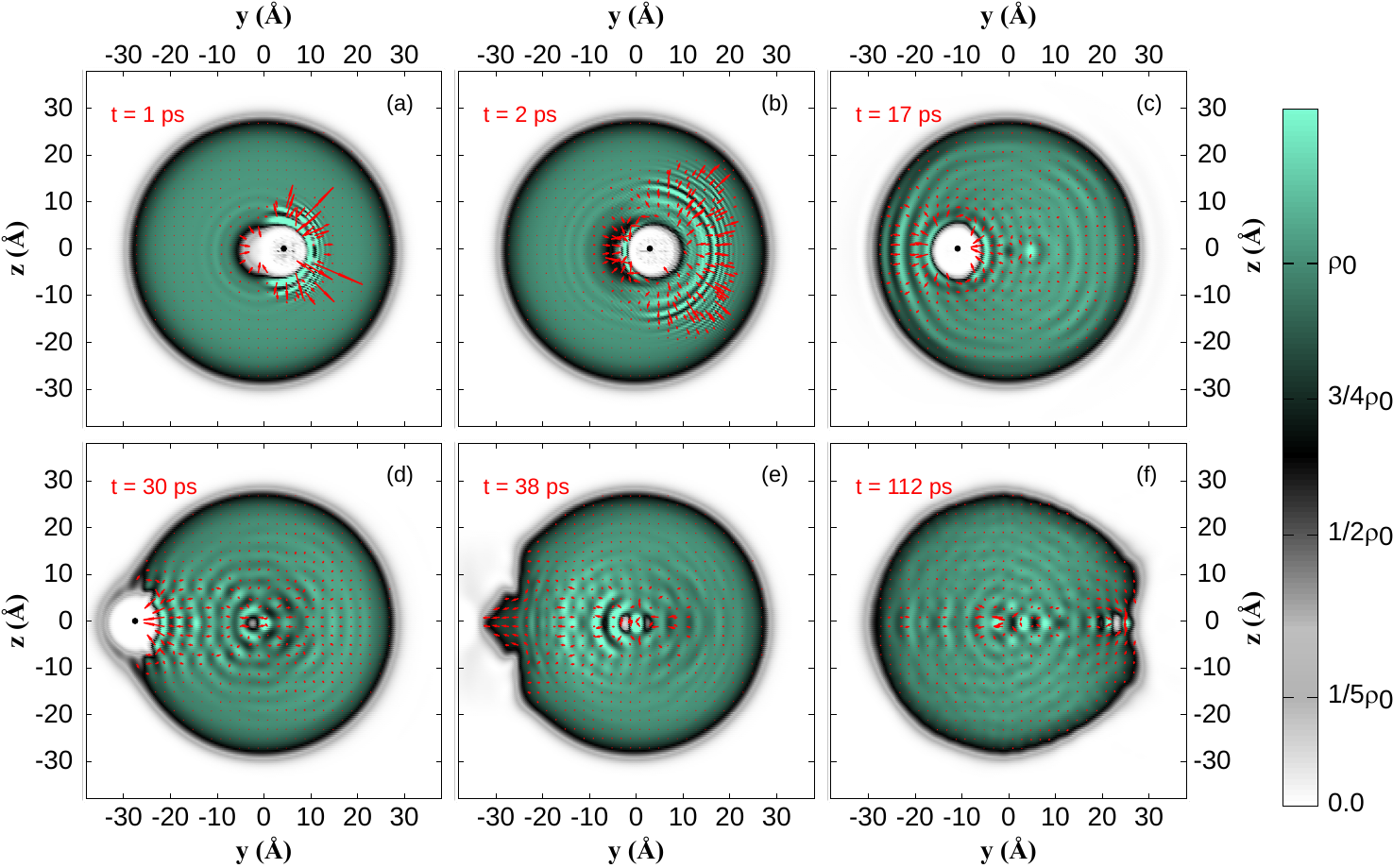}
    \caption{
        \textbf{Dynamics of a Na atom propagating through a $^4$He$_{2000}$ droplet with an initial velocity of 800~$\mps$.}
        The velocity $\vna$ is initialized along the $+y$ direction.
        Each panel shows a 2D cut of the $^4$He density in the $y$-$z$-plane. Within 1~ps, the Na atom collides with the droplet environment, which leads to a reversal of the Na atom's direction of motion.
    }\label{sfig:bulk_motion_800}
\end{figure}

We now focus on the particular case $v_{\mathrm{Na}} = 800~\mathrm{ms}^{-1}$
(brown curve in Fig.~\ref{sfig:bulk_vel_2000}). At this initial velocity, the first collision between the Na atom and the surrounding He bubble is sufficiently strong to reverse the Na atom's direction of motion.
Figure~\ref{sfig:bulk_motion_800} illustrates the trajectory of the Na
atom inside the He$_{2000}$ droplet for this case. The motion is initialized with $v_{\mathrm{Na}}$ along the $+y$-direction.
Within approximately 1~ps, the impurity collides with the bubble wall, resulting in a complete reversal of its velocity.
The Na speed decreases from 800~ms$^{-1}$ at $t = 0$~ps to zero at $t = 0.97$~ps. After the reversal, the Na atom propagates in the negative $y$-direction, see panel (b), partially recovering its speed to
$-179$~ms$^{-1}$ at $t = 1.8$~ps. Over the subsequent $\sim 5$~ps, the Na atom decelerates further to $\sim -30$~ms$^{-1}$.

From this point on, the Na atom follows a dynamical pathway analogous to that observed for the lower initial velocities. It accelerates toward the droplet surface due to the repulsive interaction with the surrounding helium, panel (c). 
The impurity bubble bursts at the surface, panel (d), generating two counter-propagating surface waves on the droplet, panel (e),
which subsequently collide and mutually annihilate, panel (f).

\clearpage 

\paragraph{Caption for Movie S1.}
\textit{File: \texttt{Mov\_01.mp4}.} 
\textbf{Dynamics of a Na atom on the surface of a $^4$He$_{1000}$ nanodroplet after it is given an initial speed of 70~$\mps$}. Animated version of Fig.~\ref{fig:surf_motion_slow}.

\paragraph{Caption for Movie S2.}
\textit{File: \texttt{Mov\_02.mp4}.} 
\textbf{Dynamics of a Na atom on the surface of a $^4$He$_{1000}$ nanodroplet after it is given an initial speed of 150~$\mps$}. Animated version of Fig.~\ref{fig:surf_motion_fast}.

\paragraph{Caption for Movie S3.}
\textit{File: \texttt{Mov\_03.mp4}.} 
\textbf{Dynamics of a Na atom propagating through a $^4$He$_{1000}$ nanodroplet after it is given an initial velocity of 150~$\mps$}. Animated version of Fig.~\ref{fig:bulk_motion}.

\end{document}